\documentclass[twocolumn]{aastex61}
\usepackage{bm}

\newcommand\starname{2MASS J151113.24--213003.0}
\newcommand\teff{T_{\rm eff}}
\newcommand\logg{\log{g}}
\newcommand\feh{[\rm{Fe}/\rm{H}]}

\newcommand\rprocess{\textit{r}-process}

\newcommand{\project}[1]{\textsl{#1}}
\newcommand{\acronym}[1]{{\small{#1}}}
\newcommand{\twomass}{\project{2MASS}}
\newcommand{\mike}{\acronym{MIKE}}
\newcommand{\apass}{\project{APASS}}
\newcommand{\wise}{\project{WISE}}
\newcommand{\spm}{\project{SPM4}}
\newcommand{\ppmxl}{\project{PPMXL}}
\newcommand{\ucac}[1]{\project{UCAC{#1}}}

\received{2017 September 1}
\revised{2017 September 12}
\accepted{2017 September 18}

\shorttitle{An EMP Star Superlatively-deficient in Sr and Ba}
\shortauthors{Casey \& Schlaufman}

\begin{document}

\title{The Universality of the Rapid Neutron-capture Process Revealed by
a Possible Disrupted Dwarf Galaxy Star\footnote{This paper
includes data gathered with the 6.5 meter Magellan Telescopes located
at Las Campanas Observatory, Chile.}}

\correspondingauthor{Andrew R. Casey}
\email{andrew.casey@monash.edu}

\author[0000-0003-0174-0564]{Andrew R. Casey}
\affiliation{
School of Physics \& Astronomy, Monash University, Clayton 3800, Victoria, Australia}
\affiliation{Faculty of Information Technology, Monash University, Clayton 3800, Victoria, Australia}

\author[0000-0001-5761-6779]{Kevin C. Schlaufman}
\affiliation{Department of Physics and Astronomy, Johns Hopkins University, 3400 N Charles St., Baltimore, MD 21218, USA}

\begin{abstract}

\noindent
The rapid neutron-capture process or \rprocess\ is thought to produce the
majority of the heavy elements ($Z>30$) in extremely metal-poor stars.
The same process is also responsible for a significant fraction of the
heavy elements in the Sun.  This universality of the \rprocess\ is one
of its characteristic features as well as one of the most important
clues to its astrophysical origin.  We report the discovery of an
extremely metal-poor field giant with $[\mathrm{Sr,Ba/H}]\approx-6.0$
and $[\mathrm{Sr,Ba/Fe}]\approx-3.0$, the lowest abundances of strontium
and barium relative to iron ever observed.  Despite its low abundances,
the star \starname\ has $[\mathrm{Sr/Ba}]=-0.11\pm0.14$ and therefore its
neutron-capture abundances are consistent with the main solar $r$-process
pattern that has $[\mathrm{Sr/Ba}]=-0.25$.  It has been suggested that
extremely low neutron-capture abundances are a characteristic of dwarf
galaxies, and we find that this star is on a highly-eccentric orbit
with apocenter $\gtrsim100$ kpc that lies in the disk of satellites
in the halo of the Milky Way.  We show that other extremely metal-poor
stars with low [Sr,Ba/H] and [Sr,Ba/Fe] plus solar [Sr/Ba] tend to have
orbits with large apocenters, consistent with a dwarf galaxy origin
for this class of object.  The nucleosynthesis event that produced the
neutron-capture elements in \starname\ must produce both strontium and
barium together in the solar ratio.  We exclude contributions from the
$s$-process in intermediate-mass AGB or fast-rotating massive metal-poor
stars, pair-instability supernovae, the weak \rprocess, and neutron-star
mergers.  We argue that the event was a Pop III or extreme Pop II
core-collapse supernova explosion.
\end{abstract}


\section{Introduction} \label{sec:intro}

The rapid neutron-capture or \rprocess\ occurs when nuclei experience
a neutron flux so intense that the lag time between neutron-capture
events is less than the corresponding $\alpha$- or $\beta$-decay
timescale \citep[e.g.,][]{bur57,cam57a,cam57b}.  Many astrophysical
events have been identified as candidates to produce conditions
favorable for the \rprocess, most (but by no means all) involving
either core-collapse supernovae or neutron-star mergers.  Chemical
evolution models of the Milky Way suggest that common events that
produce small amounts of \rprocess\ material and rare events that
produce large amounts of \rprocess\ material are both possible
\citep[e.g.,][]{qia00,arg04,mat14,she15,van15,nai17}.

While the astrophysical site of the \rprocess\ is still hotly debated,
its observational signature in the abundances of the neutron-capture
elements (i.e., $Z > 30$) has been seen from the Sun all the way
to the most ancient stars known.  Whereas the abundances of the
neutron-capture elements seems to be decoupled from the abundances
of the light (i.e., $Z \leq 13$), $\alpha$, and iron-peak (i.e.,
$21 \leq Z \leq 30$) elements in extremely metal-poor stars, no star
completely lacking in neutron-capture elements has yet been found
\citep[e.g.,][]{gil88,mcw95a,mcw95b,nor96,rya96,cay04,fra07,coh13,yon13,roe13,roe14}.

Even with the highest-quality data practically attainable, only the
abundances of a few neutron-capture elements can be measured from
high-resolution spectra of ordinary extremely metal-poor stars.  The most
frequently measurable neutron-capture elements with the strongest
atomic transitions in the visible---and the only ones observable in
neutron-capture poor stars---are strontium and barium.  In extremely
metal-poor stars, strontium ($Z = 38$ and $A = 86-88$) is representative
of the first \rprocess\ peak while barium ($Z = 56$ and $A = 134-138$) is
representative of the second peak.  Though both elements can be produced
by the slow neutron-capture or $s$-process in asymptotic giant branch
(AGB) stars, observations of metal-poor stars do not show contributions
from the $s$-process for stars with $[\mathrm{Fe/H}] \lesssim -2.8$
\citep{sim04}.

The relative abundances of the neutron-capture elements in the first
and second \rprocess\ peaks can be used to diagnose the properties and
astrophysical site of the \rprocess\ that produced the heavy elements
in metal-poor stars.  For most metal-poor stars, $[\mathrm{Sr/Ba}]$
is inversely correlated with $[\mathrm{Ba/Fe}]$.  The neutron-capture
elements in stars with $[\mathrm{Sr/Ba}] \sim 1$ have been suggested
to have been formed by the ``weak" \rprocess.  In this scenario,
the neutron flux is too weak to create the heaviest neutron-capture
elements such that the first \rprocess\ peak is produced but the
second peak is not \citep[e.g.,][]{tru02}.  On the other hand, the
neutron-capture elements in stars with $[\mathrm{Sr/Ba}] \sim -1$
were produced by an intense neutron flux that caused the \rprocess\
to seemingly skip the first peak and strongly populate the second.
Metal-poor stars with $[\mathrm{Sr/Ba}] \sim 0$ are consistent with
the ``main" \rprocess\ responsible for a significant fraction of the
neutron-capture elements in the Sun.  Interestingly, there is a hint that
in the most neutron-capture poor stars the inverse correlation between
$[\mathrm{Ba/Fe}]$ and $[\mathrm{Sr/Ba}]$ is broken and $[\mathrm{Sr/Ba}]
\sim 0$ is consistent with the main \rprocess\ \citep[e.g.,][]{fra07,lai08}.
In other words, the same main \rprocess\ seems to be responsible both for
the Sun and for the stars closest in time to the first \rprocess\ events.

It is easier to model the chemical evolution of classical and ultra-faint
dwarf (UFD) galaxies than the Milky Way, so neutron-capture abundances
in dwarf galaxies are especially useful in the effort to identify the
astrophysical site of the \rprocess.  Until recently, most observational
data suggested that at constant metallicity the metal-poor stars in UFD
galaxies had neutron-capture abundances below those of stars in the
halo of the Milky Way.  To date, neutron-capture poor giants (i.e.,
$[\mathrm{Sr,Ba/Fe}] < -1$) have been observed in the UFD
galaxies Bo\"{o}tes II \citep{ji16d}, Coma Berenices \citep{fre10}, Leo IV
\citep{sim10,fra16}, Reticulum II \citep{roe16}, Segue 1 \citep{fre14},
Segue 2 \citep{roe14b}, Triangulum II \citep{kir17}, Tucana II \citep{ji16b},
and Ursa Major II \citep{fre10}.  Such stars also occasionally occur in 
classical dwarfs like Carina \citep{ven12}, Draco \citep{ful04,coh09}, 
Hercules \citep{koc14}, and Sculptor \citep{sta13,jab15,sim15a}.  These
observations lead to the suggestion that low neutron-capture abundances
are a signature of dwarf galaxy chemical evolution \citep[e.g.,][]{fre15}.
This idea is consistent with the hypothesis that UFD galaxies are the
closest known objects to the first galaxies, possibly hosting stars
without any neutron-capture elements at all.

The recent discovery of the UFD galaxy Reticulum II has
provided another important clue to the origin of the \rprocess\
\citep{kop15a,kop15b,bec15,sim15b,wal15}.  Most stars in Reticulum II are
extremely enriched in neutron-capture elements \citep{ji16a,ji16c,roe16}.
The extreme enrichment disfavors \rprocess\ nucleosynthesis in small
amounts in individual supernovae and instead favors the injection
of a large amount of neutron-capture elements in a rare event like a
magnetorotationally driven supernova, neutron star--neutron star merger,
or black hole--neutron star merger.  Given the relative isolation
and chemically-primitive nature of Reticulum II, it provides the first
unambiguous evidence for \rprocess\ nucleosynthesis in a rare event that
produced a large amount of neutron-capture elements.

All currently known dwarf galaxies are more than 20 kpc from the Sun
though, so even metal-poor giants are usually too faint to collect
high-resolution spectra with a sufficient signal-to-noise ratio (S/N) to
measure the abundances of strontium and barium in neutron-capture
poor stars.  Much more often it is only possible to set upper limits.
While upper limits on neutron-capture abundances for dwarf galaxy stars
are consistent with the lowest neutron-capture abundances seen in the
field \citep[e.g.,][]{roe13}, in dwarf galaxies it is impossible to
differentiate stars with no neutron-capture elements from stars with
little neutron-capture elements.  That test could only be done with
brighter, nearby extremely metal-poor giants that are also neutron-capture
poor.

Here we report the discovery of \starname, an extremely metal-poor
star with the lowest abundance of neutron-capture elements
relative to iron ever observed.  We describe our observations in
Section \ref{sec:observations}, and the analysis of those data in
Section~\ref{sec:analysis}.  In Section \ref{sec:discussion} we discuss
the origin of \starname\ and the mechanisms that could have produced
the chemical abundance pattern we observe.  We conclude in Section
\ref{sec:conclusions}.

\section{Observations} \label{sec:observations}

We identified \starname\ as a candidate metal-poor star because it
satisfies criteria (1)--(4) of the photometric selection described in
\citet{sch14}.\footnote{We note that \starname\ also satisfies the full
selection as defined in \citet{sch14}.}  We initially observed the star
on 22 June 2014---the first night of Magellan observations for the pilot
``Best and Brightest" program---using the Magellan Inamori Kyocera
Echelle (\mike) spectrograph on the Magellan Clay Telescope at Las
Campanas Observatory \citep{ber03,she03}.  Our moderate signal-to-noise
ratio ($\mathrm{S/N} \approx 30$) high-resolution ($\mathcal{R} \approx
28,\!000$) snapshot spectrum confirmed that \starname\ is an extremely
metal-poor giant.  A preliminary detailed chemical abundance analysis
revealed remarkably low levels of strontium and barium plus no detectable
quantities of other neutron-capture elements.

This peculiar chemical pattern prompted us to acquire a high-quality
follow-up spectrum.  We therefore observed \starname\ with MIKE a
second time during evening twilight of 1 August 2015.  The analysis
presented here is based on those observations.  We used the 0.35\arcsec\
slit, which provides a spectral resolution of $\mathcal{R} \approx
83,\!000$ in the blue and $\mathcal{R} \approx 65,\!000$ in the red.
We obtained three exposures of 1,200 seconds each at $\mathrm{airmass}
< 1.03$ in clear conditions with variable seeing between $0.5\arcsec$
and $1.7\arcsec$.  Calibration frames were taken in the afternoon
(e.g., biases, quartz and ``milky" flat fields, ThAr lamp frames).
The signal-to-noise ratio of the stacked spectrum reaches $\mathrm{S/N}
\approx 130~\mathrm{pixel}^{-1}$ at 345 nm on the blue side, and
$\mathrm{S/N} \approx 280~\mathrm{pixel}^{-1}$ at 500 nm on the red side.

\section{Analysis} \label{sec:analysis}

We reduced the raw spectra and calibration frames using the
\href{http://code.obs.carnegiescience.edu/mike}{\texttt{CarPy}} software
package \citep{kel03,kel14}.  The first step in our analysis of the
extracted 1D spectrum was to place it in the rest frame.  We used a
spline function to continuum normalize the echelle order that includes
the \ion{Ca}{2} near-infrared triplet.  We then measured the radial
velocity by cross correlating the normalized order with a rest-frame
normalized spectrum of the metal-poor giant star HD 122563.  Using this
line-of-sight velocity, we shifted all echelle orders to the rest frame
without resampling.  After applying the heliocentric correction\footnote{In
\citet{sch14} we incorrectly reported the heliocentric radial velocity
of \starname\ as $31.0$~km~s$^{-1}$.  During the course of this work we
discovered that the radial velocity measurements reported in \citet{sch14}
did not include a correction for heliocentric motion.  An erratum is in
preparation.  The heliocentric radial velocity measured on 2014 June 22 
is $13.4 \pm 1.0$~km~s$^{-1}$.}, we find
a radial velocity of $v_{\mathrm{rad}} = 14.4 \pm 1.0$~km~s$^{-1}$.

\subsection{Stellar Parameters} \label{sec:analysis-stellar-parameters}

It is well established that the spectroscopic stellar parameters of
metal-poor giant stars suffer from systematic uncertainties due in part
to the violation of the assumptions of local thermodynamic equilibrium
\citep[LTE; e.g.,][]{kor03}.  These effects are particularly noticeable
in the classical excitation/ionization balance approach, which results
in significant biases in effective temperature $\teff$, surface gravity
$\logg$, and metallicity $\feh$ \citep[e.g.,][]{fre13a}.  For this reason,
we performed a complementary determination of the spectroscopic stellar
parameters of \starname\ to minimize the effect of systematic uncertainty.

The complementary approach makes use of the wings of Balmer lines
present in our spectrum, as Balmer line wings are extremely sensitive to
$\teff$ given a high-S/N, high-resolution spectrum.  Because they have
little degeneracy with other astrophysical parameters, the wings of
Balmer lines are arguably the most accurate way of spectroscopically
determining $\teff$ when a high-quality spectrum is available
\citep[e.g.,][]{fuh98,bar03}.  A limitation is that the profile wings
are strongly degenerate with the continuum, and to a lesser extent,
the residual line-of-sight velocity and the instrument resolution
\citep{kor02}.

In order to account for the correlations between astrophysical and
nuisance parameters, we constructed a generative model for the data.  We
used the grids of continuum-normalized hydrogen line profiles computed by
\citet{bar03}.\footnote{\href{http://www.astro.uu.se/~barklem/}{http://www.astro.uu.se/$\sim$barklem/}}
However, our tests revealed that interpolating fluxes
between grid points---even with high-order polynomial or spline
interpolation---produced systematic offsets in flux at the order of
1.0\%.  The bias direction and magnitude varied depending on position
in the Hertzprung-Russell diagram.  While this is a small offset, the
sensitivity of the Balmer line wings implies that this could propagate
to a bias in $\teff$ of about 50 K. Moreover, the number of pixels used
when comparing the model and the data here is relatively small, implying
that single-pixel systematics introduced by interpolation could have a
noticeable effect on the inferred stellar parameters.

Consequently, we chose not to interpolate the grid of pre-computed
Balmer-line profiles.  Instead we calculated the likelihood at each grid
point and marginalized over the nuisance parameters described above.
Specifically, for each pre-computed Balmer line profile we modeled the
continuum by multiplying the continuum-normalized flux with a third-order
polynomial with coefficients $\{c_m\}_{m=1}^{M}$, and we included a term
for the residual radial velocity $v_{\mathrm{res}}$.

We constructed a mask to exclude the core of each line, as well
as any stellar or telluric absorption in the neighboring wings.
For each Balmer line model we optimized the nuisance parameters by
minimizing the negative log-likelihood $-\log\mathcal{L}$.  We then
used the Gaussian approximation of a $K$-dimensional integral
$(2\pi)^{K/2}/\sqrt{\det\boldsymbol\Sigma}$ (based on the formal
covariance matrix $\boldsymbol\Sigma$) to marginalize over the nuisance
parameters at every grid point.

\begin{figure*}[ht]
\includegraphics[width=0.95\textwidth]{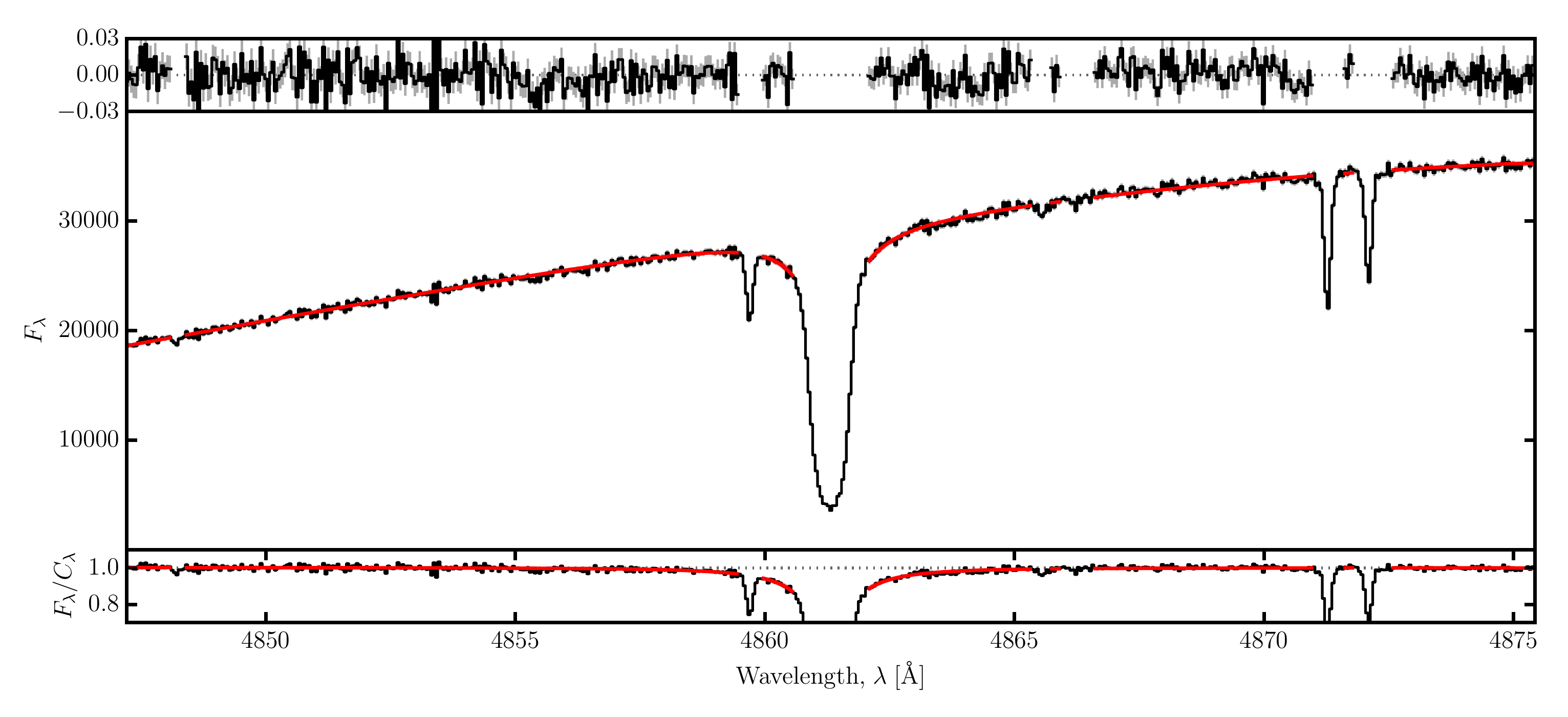}
\caption{Magellan/\mike\ spectrum of \starname\ surrounding the H-$\beta$
line.  The center panel is the spectrum itself (where the flux have
arbitrary units), the bottom panel is the spectrum divided by the 
inferred continuum, and the top panel is the residual difference between
the data and the model.  We also plot in red the optimized model closest
grid point to the maximum a posteriori solution.\label{fig01:hbeta}}
\end{figure*}

We tested this approach by fitting the H-$\alpha$, H-$\beta$,
H-$\gamma$, and H-$\delta$ lines separately using unnormalized echelle
orders of the well-studied metal-poor stars HD 122563 and HD 140283.
For these stars, we found the H-$\beta$ line to be the best indicator
of $\teff$.\footnote{A similar process leading to the same conclusion
can be found in \citet{bar02}.}  For HD 122563, our marginalized 1D
posterior peaks at $\teff = 4543^{+90}_{-130}$ K.  For HD 140283, the
posterior peaks at $\teff = 5448^{+80}_{-70}$ K.  These results are
in excellent agreement with bolometric temperatures determined using
interferometric radii \citep{hei15}.  That approach suggests that HD
122563 has $\teff = 4587\,\pm\,60$ K, a central value just 44 K hotter
than our measurement.  A similar calculation indicates that HD 140283
has $\teff = 5522\,\pm\,105$~K, 74 K warmer than our central value.
In both cases, the temperatures we find for HD 122563 and HD 140283
are within one standard deviation of either the bolometric or the
spectroscopic uncertainty.  Confident in our approach, we report the
effective temperature of \starname\ as $\teff = 4602^{+199}_{-156}$ K
(Figure \ref{fig01:hbeta}).  We note that the posteriors on $\logg$ and
[M/H] from fitting the H-$\beta$ line were uninformative: they suggested
\starname\ is more likely a metal-poor giant than a metal-rich dwarf,
but the probability distribution function is flat.

We determined $\logg$, $\feh$, and microturbulence $\xi$ by measuring
the strengths of \ion{Fe}{1} and \ion{Fe}{2} atomic absorption lines.
We normalized individual rest-frame echelle orders by fitting a spline
function to continuum regions, then resampled and stacked individual
orders onto a common dispersion array to produce a contiguous rest-frame
normalized spectrum from 331~nm to 916~nm.  Following the algorithm
described in \citet{cas14b}, we measured the strengths of individual
absorption lines by fitting Gaussian profiles.  All profiles were
scrutinized to identify and discard obviously spurious measurements.
We take the strengths and associated atomic data for all transitions
from \citet{roe10}.  We report both our measurements and the atomic data
we used in Table~\ref{tab03:line-list}.

Given our $\teff$ inferred from the wings of the H-$\beta$ line,
we performed an ionization balance to determine $\logg$, $\feh$,
and $\xi$.  We used the $\alpha$-enhanced model atmospheres from
\citet{cas04} and the 2011 version of \texttt{MOOG} \citep{sne73,sob11}
to calculate line abundances.  With $\teff = 4602$~K fixed, we
find $\logg = 0.84$, $\feh = -3.05$, and $\xi = 2.22$~km~s$^{-1}$.
However, because the surface gravities of metal-poor giant stars are
routinely underestimated due to non-LTE effects, we chose to adopt
$\logg = 1.02$ from a 10 Gyr $\alpha$-enhanced Dartmouth isochrone of
$\feh = -2.5$ \citep{dot08}.  This produced just a 0.02 dex decrease
in the mean metallicity to $\feh = -3.07$ and a 0.05 km s$^{-1}$
decrease in microturbulence to $\xi = 2.17$ km~s$^{-1}$.  It also
caused $(\langle$\ion{Fe}{1}$\rangle-\langle$\ion{Fe}{2}$\rangle)$ to
decrease to $-0.07$~dex.  We report the adopted parameters and associated
uncertainties in Table \ref{tab01:summary-parameters}.  These values
are in good agreement with the stellar parameters derived from our
reconnaissance spectrum and reported in \citet{sch14}.  The effective
temperature inferred from the H-$\beta$ line is 115 K cooler than the
\citet{sch14} measurement determined solely by excitation balance,
and the 0.1 dex changes in $\log{g}$ and [Fe/H] are consistent with the
change in inferred temperature.  These differences are within the quoted
1-$\sigma$ uncertainties of both \citet{sch14} and this study.

\begin{deluxetable}{lR}
\tablecaption{Parameters of \starname \label{tab01:summary-parameters}}
\tablehead{\colhead{Property} & \colhead{Value}}
\startdata
Right ascension $\alpha$ (J2000)                                                                       & $15~11~13.24$    \\
Declination $\delta$ (J2000)                                                                           & $-21~30~03.0$    \\
$V$-band apparent magnitude\tablenotemark{a}                                                        & 12.66 \pm 0.06 \\
$J-K_s$ color\tablenotemark{b}                                                                          & 0.62 \\
Reddening\tablenotemark{c} $E(B-V)$                                                           & 0.099 \\
\hline
Effective temperature $\teff$                                                                          & 4602^{+199}_{-156}~{\rm K}\\
Surface gravity $\logg$                                                                                        & 1.02^{+0.15}_{-0.15} \\
Metallicity $\feh$                                                                                             & -3.07 \pm 0.10 \\
Microturbulence $\xi$                                                                                          & 2.17 \pm 0.10~{\rm km~s}^{-1} \\
\hline
Radial velocity $v_{\mathrm{rad}}$                                                                                      & 14.4 \pm 1.0~{\rm km~s}^{-1} \\
Proper motion\tablenotemark{d} in $\alpha$ ($\mu_\alpha\cos\delta$)      & -4.1 \pm 0.9~{\rm mas~yr}^{-1} \\
Proper motion\tablenotemark{d} in $\delta$  ($\mu_\delta$)                       & -21.0 \pm 0.9~{\rm mas~yr}^{-1} \\
\hline
\textbf{Informed analysis (preferred)} & \\
Heliocentric distance $d_\odot$                                                                & 7.1^{+0.6}_{-0.5}~{\rm kpc} \\
Galactocentric distance $d_{\rm GC}$                                                           & 4.7^{+0.1}_{-0.1}~{\rm kpc} \\
Max height above Galactic plane $z_{\mathrm{max}}$                                                              & 106.6^{+22.8}_{-29.1}~{\rm kpc} \\
Apocenter $r_{\rm apo}$                                                                                        & 125.2^{+31.7}_{-35.5}~{\rm kpc} \\
Pericenter $r_{\rm peri}$                                                                                      & 4.3^{+0.2}_{-0.1}~{\rm kpc} \\
Eccentricity $e$                                                                                                       & 0.93^{+0.01}_{-0.02} \\
\hline
\textbf{Uninformed analysis} & \\
Heliocentric distance $d_\odot$                                                                & 4.2^{+1.0}_{-1.0}~{\rm kpc} \\
Galactocentric distance $d_{\rm GC}$                                                           & 5.2^{+0.5}_{-0.4}~{\rm kpc} \\
Max height above Galactic plane $z$ $z_{\mathrm{max}}$                                                              & 7.7^{7.9}_{-1.8}~{\rm kpc} \\
Apocenter $r_{\rm apo}$                                                                                        & 9.7^{+12.3}_{-2.8}~{\rm kpc} \\
Pericenter $r_{\rm peri}$                                                                                      & 4.0^{+0.3}_{-0.2}~{\rm kpc} \\
Eccentricity $e$                                                                                                       & 0.53^{+0.24}_{-0.14} \\
\enddata
\tablenotetext{a}{Photometry from \apass\ DR9 \citep{hen15}}
\tablenotetext{b}{Photometry from \twomass\ \citep{skr06}}
\tablenotetext{c}{Using the \citet{sch98} dust maps as updated by
\citet{sch11}}
\tablenotetext{d}{Proper motions from \ucac{5} \citep{zac17}}
\end{deluxetable}

\subsection{Detailed Chemical Abundances} \label{sec:analysis-chemical-abundances}

We report the detailed chemical abundances of \starname\ in Table
\ref{tab02:detailed-abundances}.  For most elements, we determined
individual chemical abundances from measured equivalent widths of
unblended atomic lines.  However, we adopted a synthesis approach for
molecular features (e.g., CH) and for atomic transitions with strong
isotopic splitting or hyperfine structure (specifically scandium,
manganese, cobalt, and copper).  We used molecular data for CH
from \citet{mas14} as well as hyperfine/isotopic splitting data from
\citet{kur95} for the iron-peak elements, \citet{bie99} for barium, and
\citet{law01a,law01b} for europium and lanthanum.  We assume \rprocess\
only isotopic fractions from \citet{sne08}.

\begin{deluxetable}{lrRrRR}
\tabletypesize{\scriptsize}
\tablecaption{Chemical abundances of \starname \label{tab02:detailed-abundances}}
\tablecolumns{6}
\tablehead{
\colhead{Species} &
\colhead{N} &
\colhead{$\log_\epsilon({\rm X})$} &
\colhead{$\sigma$\tablenotemark{1}} &
\colhead{[X/H]} &
\colhead{[X/\ion{Fe}{1}]}
}
\startdata
\ion{Li}{1}   &    1 &   <-0.25  &    \nodata &     <-1.30 &     <1.78 \\
C (CH)        &      &     4.80  &       0.10 &      -3.63 &     -0.55 \\
\ion{O}{1}    &    1 &    <6.92  &    \nodata &     <-1.77 &     <1.31 \\
\ion{Na}{1}   &    1 &     3.63  &    \nodata &      -2.61 &      0.47 \\
\ion{Mg}{1}   &   11 &     5.08  &       0.09 &      -2.52 &      0.55 \\
\ion{Al}{1}   &    1 &     2.79  &    \nodata &      -3.67 &     -0.59 \\
\ion{Si}{1}   &    2 &     5.09  &       0.13 &      -2.42 &      0.66 \\
\ion{K}{1}    &    1 &     2.19  &    \nodata &      -2.84 &      0.24 \\
\ion{Ca}{1}   &   17 &     3.36  &       0.06 &      -2.98 &      0.10 \\
\ion{Sc}{1}   &    6 &    -0.05  &       0.07 &      -3.20 &     -0.13 \\
\ion{Sc}{2}   &    7 &    -0.01  &       0.12 &      -3.16 &     -0.09 \\
\ion{Ti}{2}   &   36 &     1.77  &       0.13 &      -3.18 &     -0.11 \\
\ion{V}{2}    &    1 &     1.10  &    \nodata &      -2.83 &      0.25 \\
\ion{Cr}{1}   &   13 &     2.16  &       0.17 &      -3.48 &     -0.40 \\
\ion{Cr}{2}   &    2 &     2.63  &       0.02 &      -3.01 &      0.06 \\
\ion{Mn}{1}   &    7 &     1.74  &       0.09 &      -3.69 &     -0.62 \\
\ion{Mn}{2}   &    1 &     1.76  &    \nodata &      -3.67 &     -0.59 \\
\ion{Fe}{1}   &  223 &     4.42  &       0.12 &      -3.08 &      0.00 \\
\ion{Fe}{2}   &   23 &     4.49  &       0.12 &      -3.01 &      0.07 \\
\ion{Co}{1}   &    1 &    <2.91  &    \nodata &     <-2.08 &     <1.00 \\
\ion{Ni}{1}   &    9 &     3.12  &       0.12 &      -3.10 &     -0.02 \\
\ion{Cu}{1}   &    1 &    <1.36  &    \nodata &     <-2.83 &     <0.25 \\
\ion{Zn}{1}   &    2 &     1.51  &       0.01 &      -3.05 &      0.03 \\
\ion{Ga}{1}   &    1 &    <0.74  &    \nodata &     <-2.30 &     <0.77 \\
\ion{Rb}{1}   &    1 &    <1.26  &    \nodata &     <-1.26 &     <1.82 \\
\ion{Sr}{2}   &    2 &    -3.04  &       0.17 &      -5.91 &     -2.84 \\
\ion{Y}{2}    &    1 &   <-1.62  &    \nodata &     <-3.83 &    <-0.76 \\
\ion{Zr}{2}   &    1 &   <-1.29  &    \nodata &     <-3.87 &    <-0.80 \\
\ion{Nb}{2}   &    1 &   <-0.37  &    \nodata &     <-1.83 &     <1.25 \\
\ion{Mo}{1}   &    1 &   <-0.70  &    \nodata &     <-2.58 &     <0.50 \\
\ion{Sn}{1}   &    1 &    <0.71  &    \nodata &     <-1.33 &     <1.75 \\
\ion{Ba}{2}   &    1 &    -3.62  &    \nodata &      -5.80 &     -2.72 \\
\ion{La}{2}   &    1 &   <-2.68  &    \nodata &     <-3.78 &    <-0.70 \\
\ion{Ce}{2}   &    1 &   <-1.25  &    \nodata &     <-2.83 &     <0.25 \\
\ion{Pr}{2}   &    1 &   <-1.36  &    \nodata &     <-2.08 &     <1.00 \\
\ion{Nd}{2}   &    1 &   <-1.91  &    \nodata &     <-3.33 &    <-0.25 \\
\ion{Sm}{2}   &    1 &   <-1.62  &    \nodata &     <-2.58 &     <0.50 \\
\ion{Eu}{2}   &    1 &   <-2.98  &    \nodata &     <-3.50 &    <-0.42 \\
\ion{Gd}{2}   &    1 &   <-2.01  &    \nodata &     <-3.08 &     <0.00 \\
\ion{Tb}{2}   &    1 &   <-2.28  &    \nodata &     <-2.58 &     <0.50 \\
\ion{Dy}{2}   &    1 &   <-1.98  &    \nodata &     <-3.08 &     <0.00 \\
\ion{Tm}{2}   &    1 &   <-1.98  &    \nodata &     <-2.08 &     <1.00 \\
\ion{Yb}{2}   &    1 &   <-3.24  &    \nodata &     <-4.08 &    <-1.00 \\
\ion{Hf}{2}   &    1 &   <-1.48  &    \nodata &     <-2.33 &     <0.75 \\
\ion{Ir}{1}   &    1 &   <-0.45  &    \nodata &     <-1.83 &     <1.25 \\
\ion{Pb}{1}   &    1 &   <-0.83  &    \nodata &     <-2.58 &     <0.50 \\
\enddata
\tablenotetext{1}{This is the statistical uncertainty (standard deviation)
of the line abundance for each species. We estimate the systematic 
uncertainty at 0.10~dex for all elements, which we recommend be added in 
quadrature.}
\end{deluxetable}

For some elements we were only able to measure the abundance from
a single atomic transition.  One example is \ion{K}{1}, where our
abundance is calculated solely from the 767~nm line because the 770~nm
line was contaminated by strong telluric absorption.  Similarly, only
upper limits were possible for many of the elemental abundances.  This
includes lithium, oxygen, cobalt, copper, and all of the neutron-capture
elements except strontium and barium.  Upper limits were calculated from
the strongest absorption line available for each species.  Most of our
derived abundance limits are not informative despite the high quality
of our spectrum.

\subsection{Dynamics} \label{sec:analysis-dynamics}

We combined the stellar parameters inferred in Section
\ref{sec:analysis-stellar-parameters} with isochrones and multi-band
photometry to compute a posterior distance distribution to \starname.
We used $B$ and $V$ photometry from \apass\ DR9 \citep{hen15}; $J$,
$H$, and $K_s$ photometry from \twomass\ \citep{skr06}; and $W_1$,
$W_2$, and $W_3$ photometry from \wise\ \citep{wri10}.  Apparent
magnitudes were de-reddened using the \citet{sch98} dust maps as
updated by \citet{sch11}.  We used the \project{MIST} isochrone
grid \citep{dot16,cho16,pax11,pax13,pax15} supplied with the Python
\texttt{isochrones} package \citep{mor15}.  This leads us to an inferred
distance of $d_\odot = 7.5^{+0.6}_{-0.5}$~kpc.

After considering the available proper motions from HSOY \citep{alt17},
\ppmxl\ \citep{roes10}, \spm\ \citep{gir11}, \ucac{4} \citep{zac13}, and
\ucac{5} \citep{zac17}, we adopted the proper motions from \ucac{5}.
\ucac{5} incorporates Gaia DR1 astrometry to improve the \ucac{4}
proper motions and currently provides the best available proper motions
for bright stars not included in the Tycho-Gaia Astrometric Solution
\citep{mic15,lin16}.  As a result, \ucac{5} has the most precise
proper motions and reports $\mu_{\alpha\cos\delta} = -4.1 \pm 0.9$
mas yr$^{-1}$ and $\mu_\delta = -21.0 \pm 0.9$ mas yr$^{-1}$.  Given a
giant star with $d_{\odot} \approx 7$ kpc, a proper motion in excess of
20 mas yr$^{-1}$ implies a significant transverse velocity.  We note
that all of the proper motions catalogs we checked report similarly high
proper motions in declination.\footnote{\ucac{5} reports $\mu_{\delta}
= -21.0 \pm 0.9$ mas yr$^{-1}$ for \starname.  That value is consistent
with the HSOY ($-22.2 \pm 2.1$ mas yr$^{-1}$), \ppmxl\ ($-24.3 \pm 3.9$
mas yr$^{-1}$), \spm\ ($-22.2 \pm 1.3$ mas yr$^{-1}$), and \ucac{4}
($-21.7 \pm 1.4$ mas yr$^{-1}$) values.  Indeed, all five catalogs
agree within the 1-$\sigma$ uncertainty quoted by each catalog, and all
catalogs consistently report $\mu_\delta$ at a level exceeding 6 $\sigma$.
Because \starname\ is bright ($V = 12.66$), none of these entries are
likely to be in error due to misidentification.} In fact, the \ucac{5}
proper motions we adopt are slightly smaller in magnitude than the values
reported in other catalogs.

\begin{figure}
\includegraphics[width=0.425\textwidth]{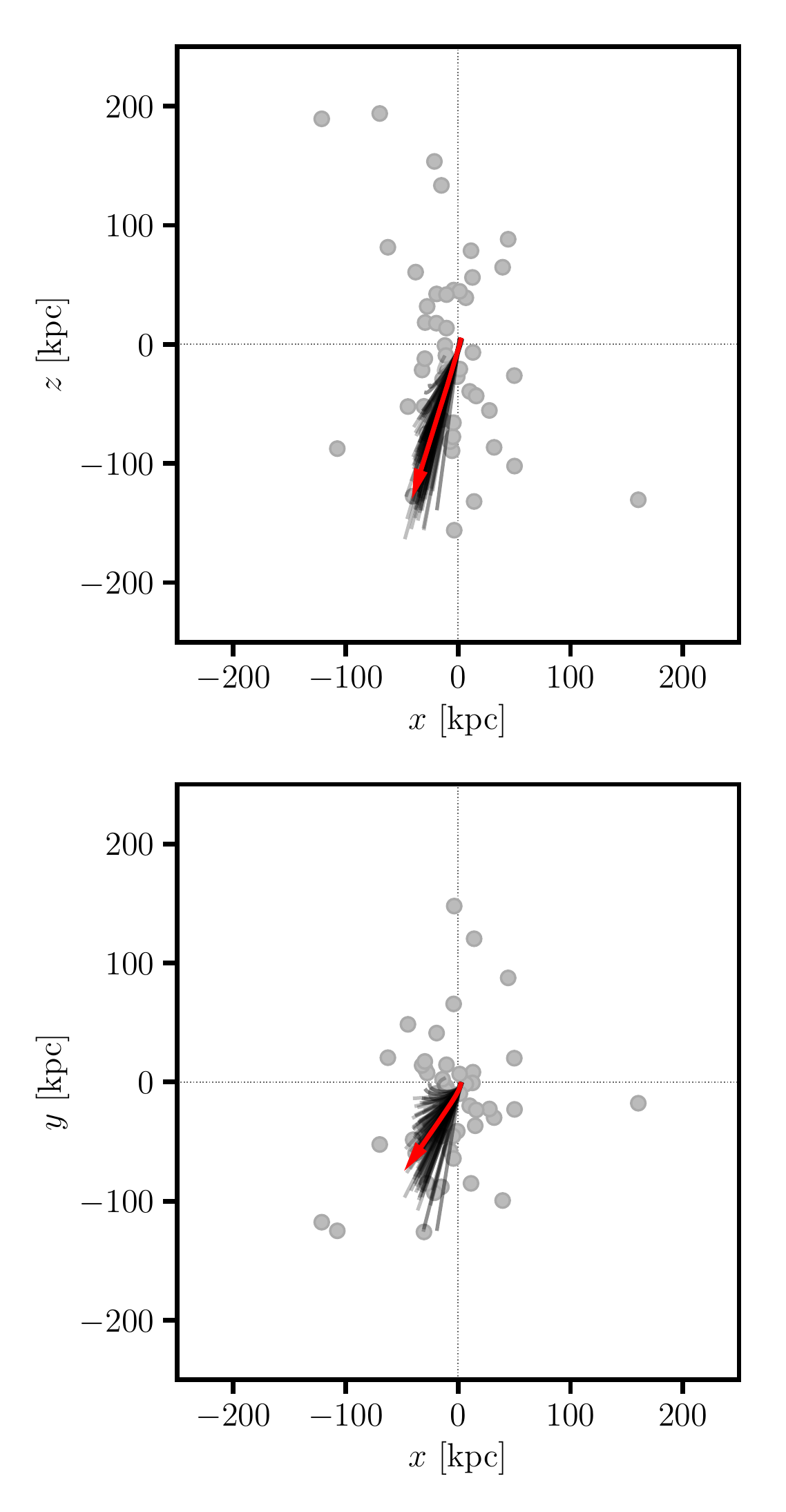}
\caption{Integrated orbits of \starname\ over 10 Gyr from 1,000 Monte
Carlo realizations in a Milky Way-like potential.  We plot individual
orbits as gray lines and the maximum a posteriori orbit as the red arrow.
We also indicate the positions of 49 classical and ultra-faint Milky
Way dwarf galaxies as gray points.  The maximum a posteriori orbit is
aligned with a plane passing through most of the Milky Way's satellite
galaxies.  While most realizations result in unbound orbits, some remain
bound.\label{fig02:orbits}}
\end{figure}

We sample 1,000 Monte Carlo realizations from the $d_\odot$
posterior and the uncertainty distributions of $v_{\mathrm{rad}}$,
$\mu_{\alpha\cos\delta}$, and $\mu_\delta$ under the assumption that
they are normally distributed.  We used each Monte Carlo realization as
initial conditions for an orbit and integrated it forward 10 Gyr in a
Milky Way-like potential using \texttt{galpy} \citep{bov15}.  We adopted
the \texttt{MWPotential2014} described by \citet{bov15}.  In that
model, the bulge is parameterized as a power-law density profile that
is exponentially cut-off at 1.9 kpc with a power-law exponent of $-1.8$.
The disk is represented by a Miyamoto-Nagai potential with a radial scale
length of 3 kpc and a vertical scale height of 280 pc \citep{miy75}.
The halo is modeled as a Navarro-Frenk-White halo with a scale length of
16 kpc \citep{nav96}.  We set the solar distance to the Galactic center
as $R_0 = 8$ kpc and the circular velocity at the Sun to $V_0 = 220$ km
s$^{-1}$ \citep{bov12}.  We find that \starname\ is approaching pericenter
($r_{\mathrm{peri}} = 4.3^{+0.2}_{-0.1}$ kpc) on a highly-eccentric orbit
($e = 0.93^{+0.01}_{-0.02}$).  We plot projections of galactic position
($x$, $y$, $z$) from all realizations in Figure \ref{fig02:orbits}.

We repeated these inferences using a less-constrained distance prior to
gauge the impact of our inferred stellar parameters.  That is, we repeated
the analysis described above with all available multi-band photometry,
but with no constraint on reddening and with increased uncertainties on
our spectroscopic stellar parameters.  We imagined that all we knew about
\starname\ was that it was an extremely metal-poor giant and set $\teff
= 4750 \pm 250$ K, $\logg = 1.5 \pm 0.5$, and $\feh = -3.0 \pm 0.5$.
We refer to this distance posterior as ``uninformed".  The net effect
is that our default ``informed" analysis places \starname\ about 3 kpc
further away from the Sun than the ``uniformed" analysis would suggest
(a 2.5-$\sigma$ difference).  Using this new distance posterior, we
integrated orbits for 10 Gyr using the same Milky Way-like potential.
The eccentricity drops from $0.93^{+0.01}_{-0.02}$ in the informed
analysis to a less precise but still significantly non-zero value of
$0.53^{+0.24}_{-0.14}$.  The change in heliocentric distance between the
two analyses makes \starname\ more likely to be bound in the uninformed
analysis, with significant differences in both $z_{\mathrm{max}}$ and
$r_{\rm apo}$.  Pericenter and Galactocentric distance remain unchanged,
and in both analyses \starname\ appears to be approaching pericenter
on an eccentric orbit.  We present the inferred orbital parameters from
both analyses in Table \ref{tab01:summary-parameters}.

\section{Discussion} \label{sec:discussion}

Our measurement of $[\mathrm{Sr,Ba/H}] \approx -6$ for \starname\
is surpassed by only three ultra-metal poor halo stars: BPS CS
22968--0014, BPS CS 22885--0096, and SMSS J031300.36--670839.3
\citep{roe14,kell14,bes15,nor17}.  The $[\mathrm{Sr,Ba/Fe}] \approx -3$
abundances we measured in \starname\ are more than a factor of two smaller
than any previous measurement for an extremely metal-poor star in the
comprehensive Stellar Abundances for Galactic Archaeology (SAGA) Database
\citep{sud08}.  Despite its record-low neutron-capture abundances,
\starname\ has $[\mathrm{Sr/Ba}] = -0.11 \pm 0.14$ and is therefore
fully consistent with the $[\mathrm{Sr/Ba}] = -0.25$ inferred for the
solar \rprocess\ \citep[e.g.,][]{sne08}.  We also showed that \starname\
is on an orbit coincident with the locations of numerous classical and
ultra-faint dwarf galaxies.  We expand on these observations and discuss
their implications in the following subsections.

\subsection{Chemical Abundances of \starname} 

\begin{figure*}
\includegraphics[width=0.95\textwidth]{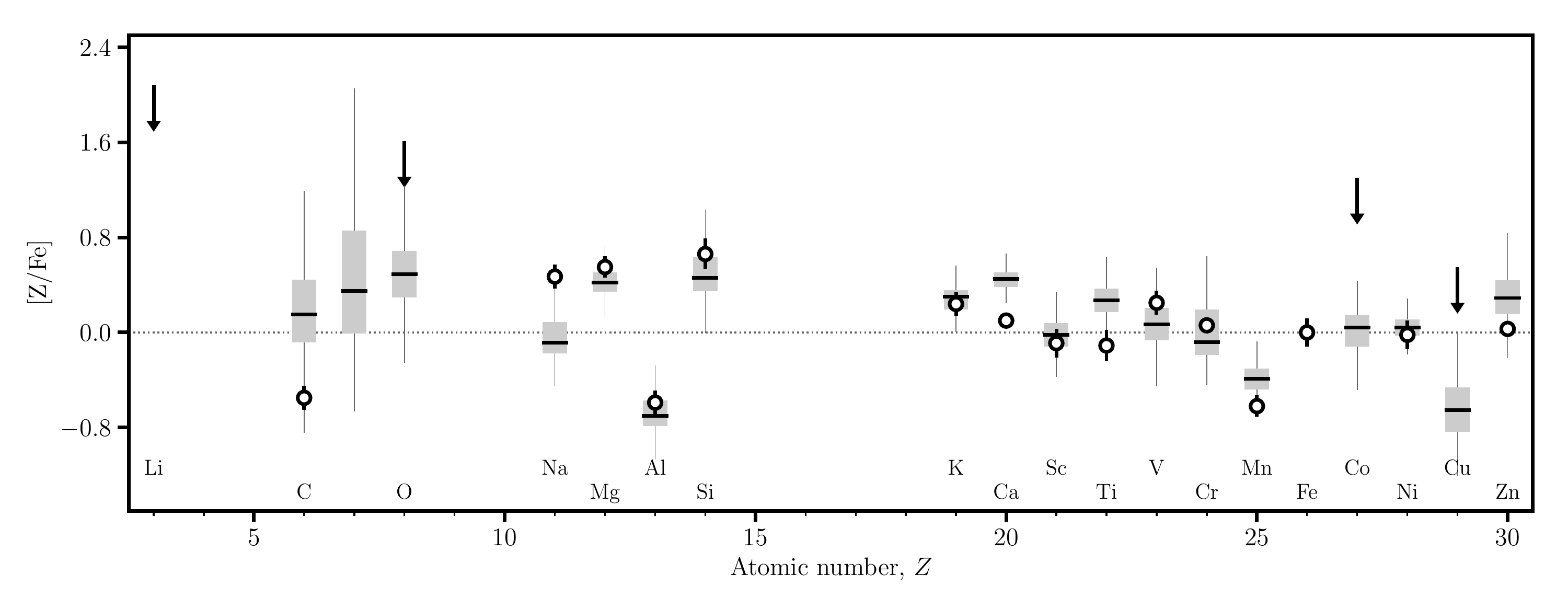}
\caption{The light, $\alpha$, and iron-peak element abundance ratios
of \starname\ (circles and upper limits) compared to the distribution
of abundance ratios from \citet{roe14} for metal-poor field giants
(boxes and whiskers).  In general, \starname\ shows element abundance
ratios that are consistent with other metal-poor field stars for $Z
\leq 30$.\label{fig03:abundances-wrt-z}}
\end{figure*}

The light, $\alpha$, and iron-peak element (i.e., $Z \leq 30$)
chemical abundance pattern of \starname\ is consistent with the
abundance ratios observed in metal-poor giants in the Milky Way (Figure
\ref{fig03:abundances-wrt-z}).  We see the largest discrepancies in the
$\alpha$ elements calcium and titanium, where we find that \starname\
has $\mathrm{[Ca,Ti/Fe]} \sim 0$.  However, two other $\alpha$ elements
magnesium and silicon are higher than the \citet{roe14} field population
($[\mathrm{Mg/Fe}] = 0.55$ and $[\mathrm{Si/Fe}] = 0.66$), producing
an average [$\langle$Mg,Ca,Si,Ti$\rangle$/Fe] value of $0.35$.  We note
that because \starname\ is a relatively cool giant, it is probably more
affected by non-LTE effects than most stars in the \citet{roe14} sample.
For this reason, we expect that some of elemental abundances of \starname\
will lie near the edge of the \citet{roe14} abundance distribution.
This effect was clearly seen by \citet{cas15} for manganese in low surface
gravity extremely metal-poor giants analyzed in a similar way.

\begin{figure}
\includegraphics[width=0.425\textwidth]{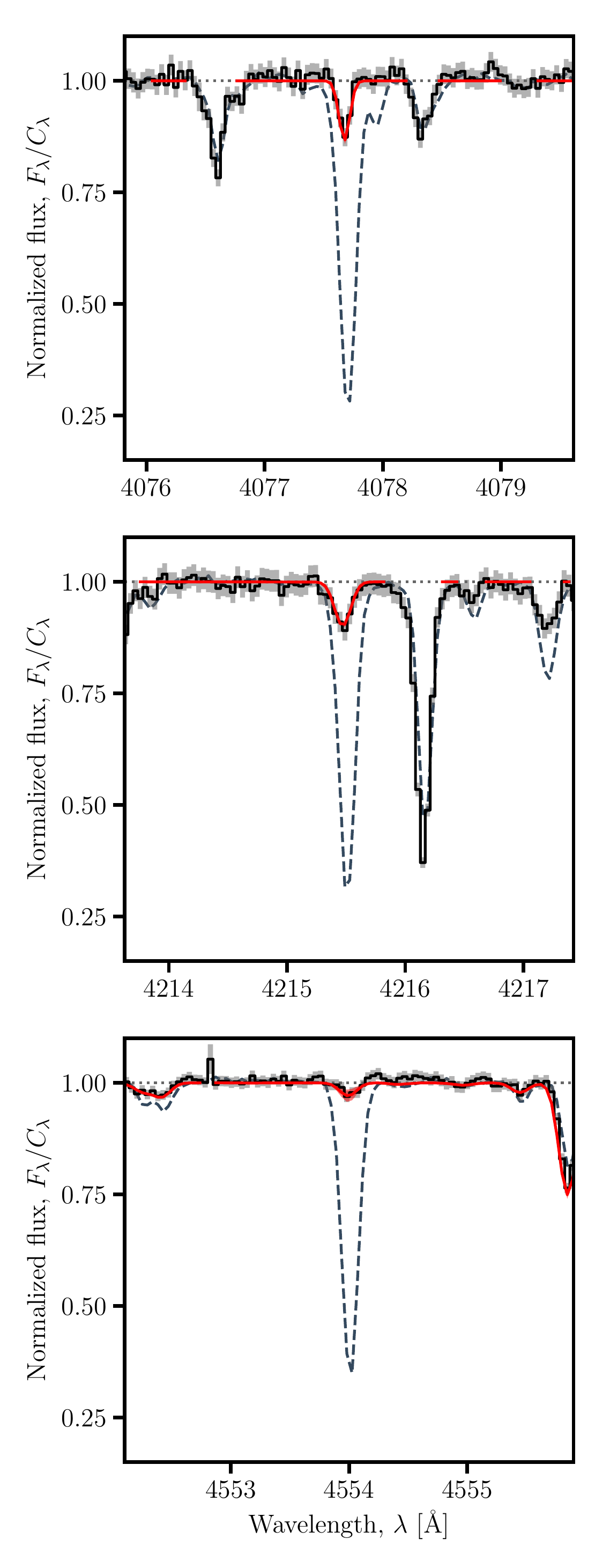}
\caption{Continuum-normalized Magellan/\mike\ spectrum surrounding the
\ion{Sr}{2} and \ion{Ba}{2} lines measurable in \starname.  We plot the
observed spectrum in black, its uncertainty in gray, and the optimized
model in red.  For comparison we also plot with the dashed line the
normalized spectrum of HD 126587, an extremely metal-poor giant with
similar stellar parameters but with solar abundance ratios of strontium
and barium.\label{fig04:ncap-spectra}}
\end{figure}

In contrast to the light, $\alpha$, and iron-peak element abundance
pattern, we find that the neutron-capture element abundance pattern in
\starname\ differs significantly from the Milky Way's halo population.
This is obvious in Figure \ref{fig04:ncap-spectra}, where we compare
the spectra of \starname\ and HD 126587---another extremely metal-poor
giant with similar stellar parameters---in the vicinity of the strongest
\ion{Sr}{2} and \ion{Ba}{2} transitions.  These transitions are usually
very strong in metal-poor giants.  However, in the spectrum of \starname\ 
we find the equivalent width of the \ion{Ba}{2} line at 455 nm is only
5m\AA.\footnote{We determined the \ion{Ba}{2} abundance by spectral 
synthesis.  We include this equivalent width measurement for completeness.} 
Similarly, we estimate that the equivalent widths of the \ion{Sr}{2} lines
at 408 nm and 422 nm to be about 15 m\AA. We note that hidden blends
may be contributing up to 9~m\AA\ to the 422 nm line, as the ratio
of oscillator strengths between these two \ion{Sr}{2} transitions indicates
that the equivalent width of the 408 nm line should be about twice as large
than the 422 nm transition. In any case, the weaknesses of these lines imply 
extremely low abundances of strontium and barium: $[\mathrm{Sr,Ba/H}]
\approx -6$ and $[\mathrm{Sr,Ba/Fe}] \approx -3$.

These extraordinary abundances are a significant departure from the
typical neutron-capture abundances seen in metal-poor giants in the halo
of the Milky Way, even after accounting for the large intrinsic scatter
seen in [Sr/Fe] and [Ba/Fe].  More quantitatively, if we represent the
abundance ratios of [Ba/Fe] seen in metal-poor giants in the halo of the
Milky Way by a Gaussian with parameters $\mu = -0.57$ and $\sigma = 0.44$
dex \citep[e.g.,][]{yon13}, then \starname\ sits more than 5 $\sigma$
below the mean value.  We find the same result for [Sr/Fe], but we note
that the two abundance ratios are correlated to some degree.

It is well known that the neutron-capture elements in the Sun are
produced in near equal proportion through the $r$- and $s$-processes
\citep[e.g.,][]{sne08}.  The $s$-process elements in the Sun are thought
to have been created in intermediate-mass AGB stars.  On the other hand,
due to the Gyr timescale required for an intermediate-mass star to
reach the AGB, there is no evidence for $s$-process contribution to the
neutron-capture abundances of stars with $[\mathrm{Fe/H}] \lesssim -2.8$
\citep[e.g.,][]{sim04}.  While there are clearly exceptions for stars
in binary systems where mass transfer appears to have occurred, this is
unlikely to be the case for \starname\ given its low neutron-capture
abundances and subsolar $[\mathrm{C/Fe}]$ ratio.  As a result, the
neutron-capture elements in \starname\ are unlikely to have been produced
in the standard $s$-process.

More exotic models of the $s$-process may be possible in the early
Universe.  \citet{ces13} used models of the $s$-process in massive,
fast-rotating, and metal-poor stars by \citet{fri12,fri16} to suggest
that these ``spinstars" could be responsible for some of the scatter
in the $[\mathrm{Sr/Ba}]$ ratio observed in stars with $[\mathrm{Fe/H}]
\lesssim -2.5$.  While the \citet{fri12,fri16} model of the $s$-process in
a $M_{\ast} = 25~M_{\odot}$ star with $\mathrm{[Fe/H]} = -7$ can explain
the $[\mathrm{Sr/Ba}] = -0.11$ ratio we observe in \starname, those
same models predict $[\mathrm{C/Fe}] \approx 1.2$ and $[\mathrm{O/Fe}]
\approx 1.4$ that are excluded by our observations.  Consequently,
the neutron-capture elements in \starname\ are unlikely to have been
produced by an exotic $s$-process in a spinstar.

\subsection{A Disrupted Dwarf Galaxy Star?} \label{sec:discuss-disrupted-ufd-star-origin}

\begin{figure}
\includegraphics[width=0.425\textwidth]{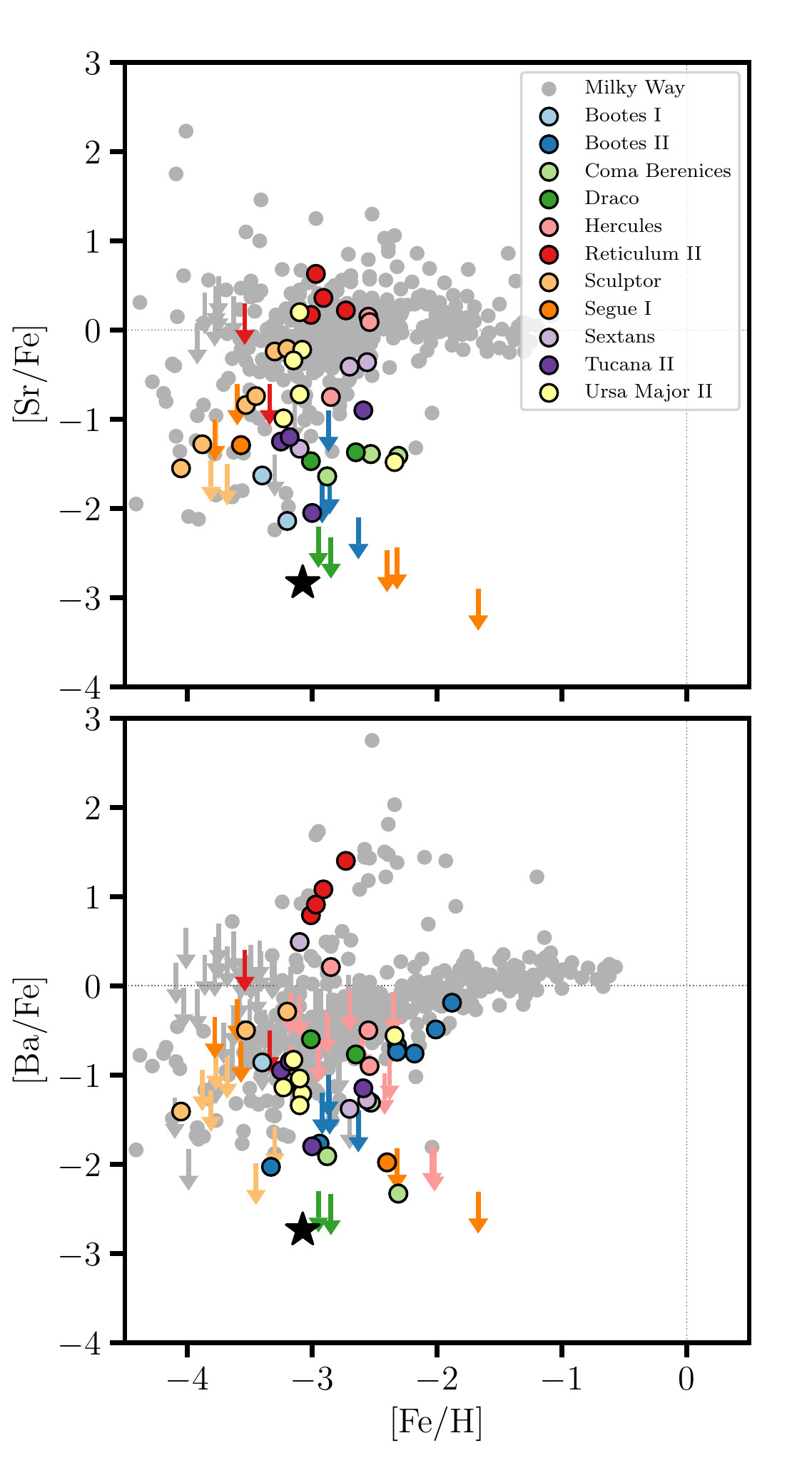}
\caption{Strontium and barium abundance ratios of giant stars in the Milky
Way field and dwarf galaxies.  We indicate the position of \starname\
with a black star in both panels.  We plot Milky Way giants from the large
samples of \citet{fra07}, \citet{ish13}, \citet{yon13}, and \citet{roe14} as gray points.  We plot stars
in classical and ultra-faint dwarf galaxies as colored points or arrows
for upper limits.  \starname\ has significantly lower neutron-capture
abundances than any star previously known.\label{fig05:n-cap}}
\end{figure}

The \rprocess-enhanced stars in Reticulum II not withstanding, it appears
that many metal-poor giants in dwarf galaxies have low abundances
of the neutron-capture elements \citep[e.g.,][]{fre15}.  To put the
abundances of \starname\ in context, we plot in Figure \ref{fig05:n-cap}
its $[\mathrm{Sr/Fe}]$ and $[\mathrm{Ba/Fe}]$ abundance ratios along
with a sample of both Milky Way halo and dwarf galaxy giant stars
\citep{ful04,koc08,koc13,coh09,coh10,fre10,fre14,nor10,sim10,taf10,
hon11,kir12,gil13,ish14,jab15,sim15a,fra16,ji16a,ji16b,ji16c,kir17}.
Our measurement of $[\mathrm{Sr,Ba/H}] \approx -6$ for \starname\
is lower than any reported measurements for dwarf galaxy members.
Our measurement of $[\mathrm{Sr,Ba/Fe}] \approx -3$ is surpassed only
for strontium by SDSS J100710.07+160623.9, a star with $[\mathrm{Fe/H}]
= -1.66$ in the UFD galaxy Segue 1.

The extraordinarily-low neutron capture abundances of \starname\ and
the apparent overabundance of such stars in dwarf galaxies relative to
the halo begs the question: did \starname\ form in an undiscovered or
now-disrupted dwarf galaxy?  If \starname\ formed in a dwarf galaxy,
then regardless of the fate of its parent system the orbital properties
of \starname\ should be very similar to those of its parent system.
In Figure \ref{fig02:orbits} we plotted possible orbits of \starname\
consistent with the best available observational data as well as the
positions of 49 classical or ultra-faint dwarf galaxies that orbit the
Milky Way.  The known dwarf galaxies are arrayed in a plane that is nearly
coincident with the orbital plane of \starname\ \citep[e.g.,][]{paw12}.
The alignment between its orbit and the disk of satellites is a hint
that \starname\ may have formed in a galaxy with a similar orbit.

Cosmological simulations preformed by \citet{wet11} suggest that some
dwarf galaxies will be accreted by Milky Way-like systems on radial orbits
with high eccentricities $\langle{}e\rangle \gtrsim 0.85$.  Those dwarf
galaxies on radial orbits are more likely to experience tidal stripping
near pericenter and be disrupted \citep[e.g.,][]{dek03,tay04,mcc08}.
Intriguingly, \starname\ is on such an orbit.  \starname\ remains on
a highly-eccentric orbit even in our uninformed analysis with extremely
conservative uncertainties on its spectroscopic properties.

The small number of giant stars observed in low-mass dwarf galaxies in
the halo implies that any tidally-disrupting low-mass dwarf galaxy at
$\sim\!\!10$ kpc would have an extremely low surface brightness and be
distributed over many tens of square degrees.  Pan-STARRS provides the
best imaging data along the line of sight to \starname, but we see no
evidence of a dwarf galaxy or surface brightness substructure in that
part of the sky.  However, given the possibility that the parent system
could be very dispersed, the absence of evidence for low-mass dwarf
galaxy or surface brightness substructure is not evidence of absence.

While the abundances and orbit of \starname\ provide circumstantial
evidence of its possible origin in a now-disrupted dwarf galaxy, it
is impossible to firmly make that association with those data alone.
To explore the possible association further, we searched the SAGA
database for other metal-poor giants with $[\mathrm{Fe/H}] \lesssim
-3.0$, $\log{g} \lesssim 3.0$, and $[\mathrm{Sr,Ba/Fe}] \lesssim -1.5$.
We cross matched those stars with the \ucac{5} proper motion database and
retained those stars with 5-$\sigma$ proper motion detections.  The five
stars so selected were BPS CS 22885--0096 and BPS CS 22960--0048 from
\citet{roe14} along with BPS CS 22968--0014, BPS CS 29502--0042, and BPS
CS 30325--0094 from \citet{fra07}.  We estimated the distance to each star
and calculated their orbits using radial velocities from \citet{bon09}
and \citet{roe14} following the same approach that we described in Section
\ref{sec:analysis-dynamics}.  We find that all five stars save BPS CS
29502--0042 have $r_{\mathrm{apo}} \gtrsim 20$ kpc and are therefore
members of the outer halo.  Assuming the $r_{\mathrm{apo}}$ distribution
from \citet{bee17}, the probability that four of five randomly-selected
extremely metal-poor field stars have $r_{\mathrm{apo}} \gtrsim 20$
kpc is only 3\% (about 1.9 $\sigma$).  The outer halo is known to have
a significantly larger fraction of its stellar population contributed by
disrupted dwarf galaxies than the inner halo \citep[e.g.,][]{car07,sch12},
so the statistical association of the neutron-capture poor phenomenon
with the outer halo also points to its association with dwarf galaxies.

We conclude this subsection by summarizing three relevant facts.  
First, \starname\ is on an eccentric orbit with apocenter $\gtrsim$100~kpc
aligned with the disk of satellites in the halo of the Milky Way.  Second, 
neutron-capture poor giants appear to be more common in dwarf galaxies 
than in the field halo population.  Third, neutron-capture poor stars in 
the halo tend to have large apocenters. While individually none of these 
observations are conclusive proof of \starname's origin in a 
now-disrupted dwarf galaxy, we argue that the preponderance of evidence 
points in that direction.  If \starname\ did originate in a dwarf galaxy,
then it would be much closer and brighter than any other star formed in a 
dwarf galaxy.  Thanks to their proximity and therefore bright apparent
magnitudes, \starname\ and the other neutron-capture poor extremely 
metal-poor giants potentially provide the best opportunity to study 
the origin of the heaviest elements in dwarf galaxies.

\subsection{Origin of the \rprocess\ material in \starname}

\starname\ has extraordinarily-low neutron capture abundances
$[\mathrm{Sr,Ba/H}] \approx -6.0$ and $[\mathrm{Sr,Ba/Fe}] \approx -3.0$,
but solar $[\mathrm{Sr/Ba}] \approx -0.1$.  Any possible origin for its
neutron-capture elements must occur promptly and be able to simultaneously
explain both observations.  We have shown that the neutron-capture elements
in \starname\ cannot be attributed to a standard or exotic $s$-process,
so they must have been created by the \rprocess.  While the origin of the
\rprocess\ has been thoroughly debated since it was first described by
\citet{bur57} and \citet{cam57a,cam57b}, we will limit our discussion
to the two sites most favored today: core-collapse supernovae and
neutron-star mergers.

For most of its history, core-collapse supernovae have been the favored
astrophysical origin of the \rprocess.  Many models in the literature can
produce neutron-capture material following a core-collapse supernovae,
and most rely on a neutrino-driven wind expected to be produced
in the first few seconds following the creation of a neutron star.
In the high temperature and entropy conditions in the evacuated bubble
surrounding the newly-formed neutron star, a neutrino can bind with a
proton to produce a neutron and a positron, driving an expanding wind
and producing a bubble of neutron-rich material to seed the \rprocess\
\citep[e.g.,][]{mey92}.  However, the \rprocess\ yields in this scenario
are extremely sensitive to the equation of state, the electron fraction,
the neutrino properties, and the $\beta$-decay rate, among other factors
\citep[e.g.,][]{qia96,hof97,tho01}.

Most models find that neutrino-driven winds from a proto-neutron
star following a core-collapse supernova can be very effective at
producing less-massive $r$-process nuclei (e.g., strontium, yttrium,
and zirconium).  In contrast, it is generally difficult to produce
heavier nuclei with $A \gtrsim 130$ like barium, and this under
production of heavy nuclei relative to the solar \rprocess\ pattern
has remained a problem \citep[e.g.,][]{rob10,rob12}.  Some groups have
succeeded in producing solar ratios of \rprocess\ material following
core-collapse supernovae, but these successes critically depend on a
sensitive time-dependent electron fraction and other uncertain factors.
They also require strong magnetic fields or neutron star masses in excess
of $2~M_{\odot}$ to create the heavier nuclei in significant proportions
\citep[e.g.,][]{wan13}.  While neutron star masses near $2~M_{\odot}$ 
have been inferred in the field \citep[e.g.,][]{lat12}, the causality 
limit breaks above 2.4 $M_\odot$.  In short, most models of core-collapse
supernovae generally have trouble producing heavy $r$-process elements 
like barium in the solar \rprocess\ ratio unless somewhat exotic 
conditions are invoked.

Population III stars with $140~M_{\odot} \lesssim M_{\ast} \lesssim
260~M_{\odot}$ could end their lives as pair-instability supernovae 
and may produce neutron-capture elements in some situations.
If the neutron-capture elements in \starname\ were the result of a
pair-instability supernova, then \citet{heg02} predicted that we should
observe a strong odd/even effect in the light and $\alpha$ elements as
well as very little zinc.  We do not see an odd/even effect and we see
$[\mathrm{Zn/Fe}] = 0.03$.  Accordingly, the neutron-capture elements
in \starname\ are unlikely to have been produced by the \rprocess\
in a pair-instability supernovae.

Nevertheless, core-collapse supernovae are an attractive astrophysical
site for the creation of the neutron-capture elements in \starname.
Supernovae must have occurred before the formation of \starname\ as
its light, $\alpha$, and iron-peak element abundances require them.
Supernovae also occur promptly, so there is no timescale problem.
If supernovae produce \rprocess\ elements, then they produce them only in
small quantities---less than $10^{-7}~M_{\odot}$ \citep[e.g.,][]{tsu00}.
An event that promptly produces light, $\alpha$, iron peak, and a small
amount of neutron capture elements is fully consistent with the $5 \times
10^{-14}~M_{\odot}$ and $2 \times 10^{-14}~M_{\odot}$ of strontium and
barium in \starname.  Indeed, the models of \citet{heg10} indicate that
Pop III supernovae in the mass range $40~M_{\odot} \lesssim M_{\ast}
\lesssim 75~M_{\odot}$ with a range of energies, piston locations,
and mixing parameters can accommodate the abundances of \starname.

Neutron-star or black hole--neutron star mergers have attracted
significant recent attention as a possible astrophysical site of the
\rprocess\ \citep[e.g.,][]{lat76,lat77,eic89,dav94,fre99}.  The escape
of neutron-rich matter from the deep potential well of a neutron star
during a merger produces rapid neutron-capture nucleosynthesis consistent
with the main solar \rprocess.  Neutron-rich material can escape the
merger in unbound tidal tails or in a wind from a rotationally-supported
accretion disk left behind by the merger and seed the interstellar medium
with neutron-capture elements.  Neutron-star or black hole--neutron star
mergers are also thought to be capable of producing elements in all three
peaks of the main solar \rprocess\ \citep[e.g.,][]{wu16,rob17,fer17}.
Because of their expected rarity compared to core-collapse supernovae,
neutron-star or black hole-neutron star mergers are thought to produce
significant quantities of neutron-capture material \citep[e.g.,][]{arg04}.
Unlike supernovae, they require a significant lag time between the onset
of star formation and injection of neutron-capture elements into the
interstellar medium.  They also do not produce significant amounts of
the light or $\alpha$ elements.

\begin{figure}
\includegraphics[width=0.45\textwidth]{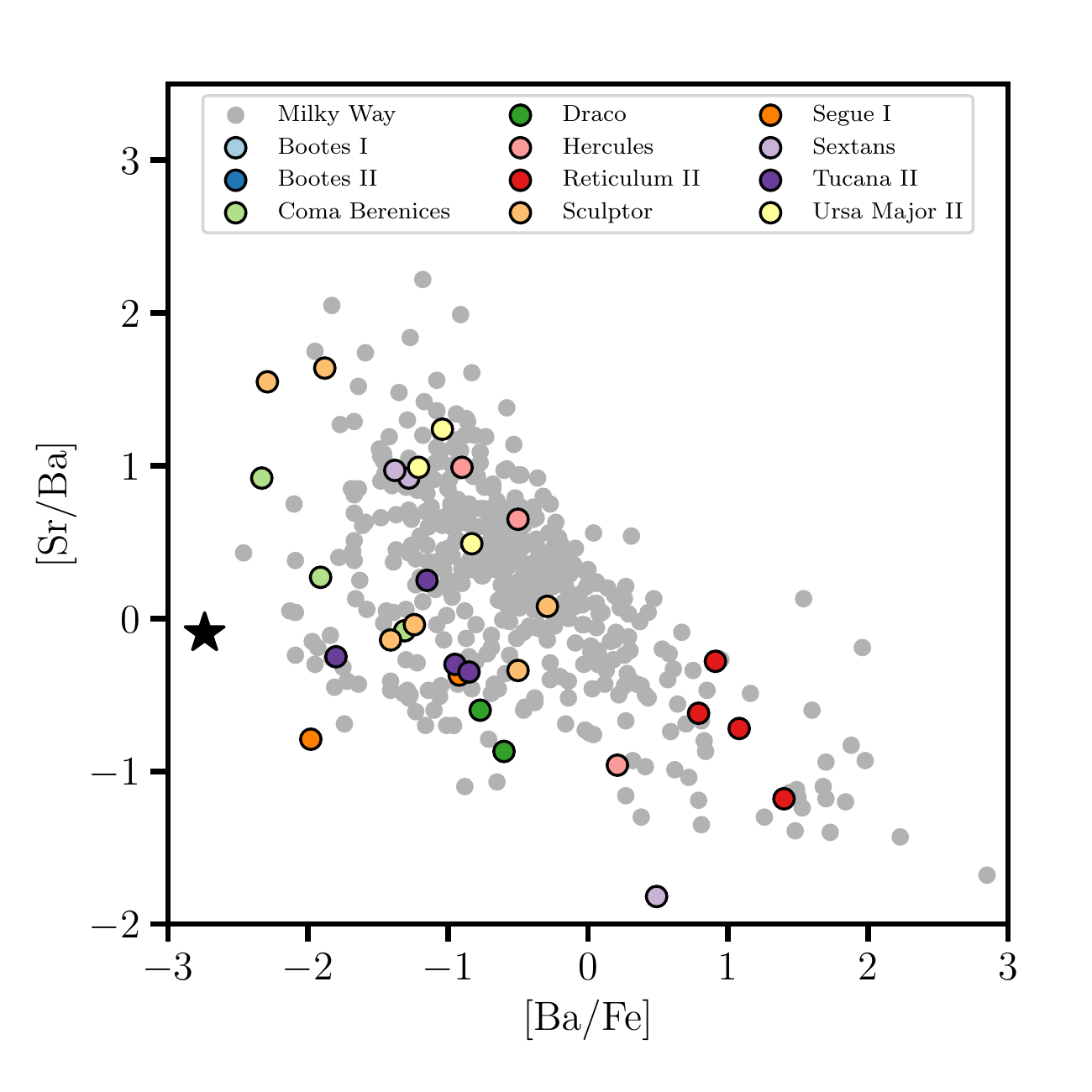}
\caption{[Ba/Fe] versus [Sr/Ba] for very metal-poor giants omitting stars
with only upper limits on strontium or barium.  We indicate the position
of \starname\ with a black star.  We plot Milky Way giants from the large
samples of \citet{fra07}, \citet{ish13}, \citet{yon13}, and \citet{roe14}
as gray points.  We plot
stars in classical and ultra-faint dwarf galaxies as colored points.
\starname\ has extremely low [Ba/Fe] but solar [Sr/Ba].\label{fig06:n-cap2}}
\end{figure}

The observation of \rprocess\ enrichment in the metal-poor giant stars in
the UFD galaxy Reticulum II by \citet{ji16a} and \citet{roe16} provides
empirical evidence for a rare, prolific source of \rprocess\ elements as
well as the nucleosynthesis produced by such an event.  Because Reticulum
II is so far unique in a sample of about 20 UFD galaxies, the process that
created its neutron-capture elements is rare.  The empirical properties of
the nucleosynthesis event in Reticulum II can be inferred from Figures
\ref{fig05:n-cap} and \ref{fig06:n-cap2}: $[\mathrm{Sr/Fe}] \approx
0.5$, $[\mathrm{Ba/Fe}] \approx 1.0$, and $-1 \lesssim [\mathrm{Sr/Ba}]
\lesssim 0$.  Given its isolation, \citet{ji16a} suggested that it would
take more than 1,000 typical core-collapse supernovae to produce the mass
of \rprocess\ elements present in Reticulum II.  This is incompatible with
the star formation history and metallicity distribution of the galaxy.
Instead, \citet{ji16a} favored an event that produced a large amount of
neutron-capture elements without creating much iron.  They suggested
that either a neutron-star merger or a magnetorotationally-driven
supernova \citep[e.g.,][]{win12} were the best candidates, though the
full distribution of neutron-capture material in Reticulum II may still
require an additional source \citep{ji16c}.

Taking all of these facts into account, the best explanation for the
origin of the neutron-capture elements in \starname\ is a Pop III or
extreme Pop II core-collapse supernova.  Core-collapse supernovae are
prompt, and at least one such explosion is necessary to have created the
$Z \leq 30$ elements in \starname.  Despite the difficulties theoretical
models of the \rprocess\ in core-collapse supernovae have producing
elements with $A \gtrsim 130$, our observation empirically suggests that
somewhere in some core-collapse supernovae strontium and barium are both
produced in the solar ratio.  Rare \rprocess\ events---presumably mergers
involving a neutron star or magnetorotationally-driven supernova---like
the event that occurred in Reticulum II produce $[\mathrm{Sr/Fe}]$ and
$[\mathrm{Ba/Fe}]$ far too high to explain our observations of \starname.
This conclusion is strengthened by the evidence we presented above that
\starname\ formed in a now-disrupted dwarf galaxy.  That is, Reticulum II
demonstrates the abundance signatures of stars in dwarf galaxies seeded
with neutron-capture elements by a rare, prolific event.  The fact that
\starname\ does not fit that pattern suggests its neutron-capture elements
were created by a qualitatively different process.

\begin{figure*}
\includegraphics[width=\textwidth]{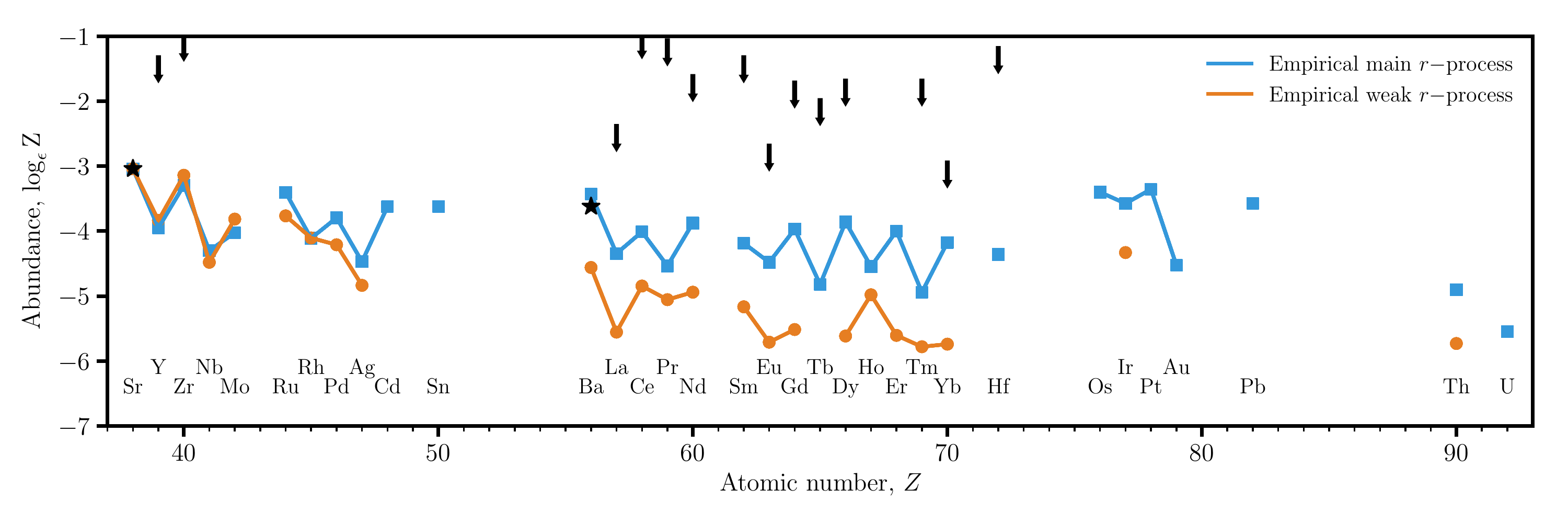}
\caption{Neutron-capture element abundances and upper limits for
\starname.  We also plot the empirical yields of the main and weak
$r$-processes in blue and orange respectively.  The $y$ axis has been
clipped for clarity and not all upper limits are shown.  We have scaled
both theoretical curves to the [Sr/H] abundance of \starname.  The [Ba/Fe]
ratio of \starname\ indicates that its neutron-capture abundances are
consistent with the main \rprocess.\label{fig07:empirical-yields}}
\end{figure*}

While \starname\ is superlative in that it has the lowest
$[\mathrm{Sr/Fe}]$ and $[\mathrm{Ba/Fe}]$ yet seen, the most
neutron-capture poor stars previously known share many of its properties.
\citet{fra07} and \citet{lai08} found that extremely metal-poor stars
in the Milky Way with decreasing $[\mathrm{Ba/H}]$ abundance ratios
showed higher $[\mathrm{Sr/Ba}]$ representative of a possible weak
\rprocess\ in contrast the solar $[\mathrm{Sr/Ba}]$ expected in the
main \rprocess.  In Figure \ref{fig07:empirical-yields}, we plot the
abundances of \starname\ along with the empirical yields of both the
main and weak \rprocess.  We follow \citet{roe17} and define the weak
\rprocess\ as the average abundances of the stars HD 88609 and HD 122563
from \citet{hon06,hon07}.  We define the main \rprocess\ as the average
abundances of the stars BPS CS 22892--0052 and BPS CS 31082--0001 from
\citet{sne03,sne09} and \citet{hil02}.  We extend to $[\mathrm{Ba/H}]
\lesssim -6$ the results of \citet{fra07} and \citet{lai08} that for
neutron-capture poor stars with $[\mathrm{Ba/H}] \lesssim -4.5$ the main
\rprocess\ is favored over the weak \rprocess\  preferred for common
extremely metal-poor stars.  \citet{fra07} went on to suggest different
regimes where weak \rprocess\ models could play a role before concluding
that it was unnecessary for stars with $[\mathrm{Ba/H}] \lesssim -4.5$.
As a result, we assert that the \rprocess\ material in \starname\ is
consistent with being produced wholly from the main \rprocess\ with no
contributions from the weak \rprocess.

These observations extend the parameter space in which the \rprocess\
appears to be universal.  The universality of the \rprocess\ usually
refers to the idea that the neutron-capture abundance pattern in
\rprocess\ enhanced metal-poor stars is very similar to the inferred
solar \rprocess\ pattern describes.  This remarkable observation implies
that the same \rprocess\ pattern observed in the Sun is seen in both
extraordinarily neutron-capture rich and extraordinarily neutron-capture
poor extremely metal-poor stars.

The extraordinarily low but non-zero abundances of neutron-capture
elements we observed in \starname\ and our conclusion that Pop III or
extreme Pop II supernovae produced neutron-capture elements leads to
the prediction that there will never be a Pop II star found without
any neutron-capture elements.  Our inference that \starname\ formed
in a now-disrupted dwarf galaxy suggests that stars in dwarf galaxies
with low upper limits on the their strontium and barium abundances
have comparable abundances to those we see in \starname\ and are not
neutron-capture free.  We predict that when high-resolution spectrographs
on 30 m class telescopes are used to investigate the neutron-capture
abundances of stars in dwarf galaxies that are currently too faint
for high-resolution spectroscopy, they will not find any stars without
neutron-capture elements.  We suggest that the same will be true in the
large number of as-of-yet undiscovered UFD galaxies that will be found
by the Large Synoptic Survey Telescope.

\section{Conclusions} \label{sec:conclusions}

We described the properties of \starname, a star we discovered and found to
have  $[\mathrm{Sr,Ba/H}] \approx -6.0$ and $[\mathrm{Sr,Ba/Fe}] \approx
-3.0$.  These are the lowest abundances of strontium and barium relative
to iron ever observed.  In addition, the star has $[\mathrm{Sr/Ba}]
= -0.11 \pm 0.14$ consistent with the scaled solar \rprocess\ pattern
that has $[\mathrm{Sr/Ba}] = -0.25$.  We find that \starname\ has an
extreme orbit taking it beyond 100 kpc that is aligned with the disk of
satellites in the halo of the Milky Way.  These observations indicate
that it may have formed in a now-disrupted dwarf galaxy.  We confirm that
other neutron-capture poor stars preferentially have Galactic orbits
with apocenters beyond 20 kpc, suggesting that neutron-capture poor stars
belong to the outer halo stellar population.  Since the contribution
of disrupted dwarf galaxies to the halo increases significantly with
distance and many stars in surviving dwarf galaxies are neutron-capture
poor, this relationship supports the idea that neutron-capture poor stars
form in dwarf galaxies.  After considering both the standard and exotic
$s$-process as well as the \rprocess\ in core-collapse supernovae, pair
instability supernovae, and mergers involving a neutron star we concluded
that the explosion of a Pop III or extreme Pop II star provided the best
explanation for the origin of the neutron-capture elements in \starname.
Despite the apparent theoretical difficulty in doing so, this empirical
inference requires that both strontium and barium be formed in the solar
ratio somewhere in some kind of core-collapse supernovae.

\acknowledgments
We thank the anonymous referee for a prompt, detailed, and insightful
review.
We thank Prashin Jethwa (Cambridge) for assistance with observations.
We thank Denis Erkal (Cambridge), Brendan Griffen (Scaled Biolabs), Alex
Ji (Carnegie Observatories), Andy Lipnicky (Rochester IoT), Andy 
McWilliam (Carnegie Observatories), Sergey Koposov
(Carnegie Mellon), David Nataf (Johns Hopkins), Adrian Price-Whelan (Princeton)
and Luke Roberts (Michigan State) for valuable discussions.  
A.~R.~C. acknowledges support from the Australian Research Council 
through Discovery Project grant DP160100637.
This research has made use of: Astropy,
a community-developed core Python package for Astronomy \citep{ast13};
NASA's Astrophysics Data System Bibliographic Services; and the SIMBAD
database and VizieR catalog access tools provided by CDS, Strasbourg,
France. The original description of the VizieR service was published
by \citet{och00}.  This publication makes use of data products from the
Two Micron All Sky Survey, which is a joint project of the University of
Massachusetts and the Infrared Processing and Analysis Center/California
Institute of Technology, funded by the National Aeronautics and Space
Administration and the National Science Foundation.  This publication
makes use of data products from the Wide-field Infrared Survey Explorer,
which is a joint project of the University of California, Los Angeles, and
the Jet Propulsion Laboratory/California Institute of Technology, funded
by the National Aeronautics and Space Administration.  This research was
made possible through the use of the AAVSO Photometric All-Sky Survey
(APASS), funded by the Robert Martin Ayers Sciences Fund.

\vspace{5mm}
\facilities{Magellan:Clay (MIKE echelle spectrograph)}

\software{
	\texttt{astropy} \citep{ast13},
    \texttt{CarPy} \citep{kel03,kel14},
    \texttt{galpy} \citep{bov15},
    \texttt{isochrones} \citep{mor15},
    \texttt{matplotlib} \citep{hun07},
    \texttt{MOOG} \citep{sne73},
    \texttt{numpy} \citep{van11},
    \texttt{scipy} \citep{oli01}
}


\clearpage
\begin{longrotatetable}
\begin{deluxetable*}{llrRRR}
\tablecaption{Individual line abundances of \starname\ and associated atomic data \label{tab03:line-list}}
\tablecolumns{6}
\tablehead{
	\colhead{Wavelength} &
	\colhead{Species} & 
	\colhead{$\chi$} & 
	\colhead{$\log{gf}$} &
	\colhead{E.W.} &
	\colhead{$\log_\epsilon({\rm X})$} \\
\colhead{[{\rm \AA}]} & & \colhead{[eV]} & & \colhead{[{\rm m\AA}]}
}
\startdata
 6707.00    & Li I  & (synth) &        &                       & <-0.28 \\
 4316.00    & C (CH)& (synth) &        &                       &   4.80 \\
 7771.94    & O I   &   9.146 &  0.324 &   {4.0}^{+2.8}_{-2.1} &  <6.92 \\
 5688.20    & Na I  &   2.100 & -0.450 &   {5.2}^{+1.3}_{-1.9} &   3.63 \\
 3829.36    & Mg I  &   2.710 & -0.208 & {155.0}^{+2.4}_{-2.3} &   5.22 \\
 3838.29    & Mg I  &   2.720 &  0.490 & {203.7}^{+2.5}_{-2.2} &   4.98 \\
 3986.75    & Mg I  &   4.346 & -1.030 &  {21.7}^{+1.7}_{-1.4} &   5.13 \\
 4057.51    & Mg I  &   4.346 & -0.890 &  {31.9}^{+2.7}_{-2.9} &   5.20 \\
 4167.27    & Mg I  &   4.346 & -0.710 &  {38.7}^{+1.3}_{-1.5} &   5.13 \\
 4571.10    & Mg I  &   0.000 & -5.688 &  {53.6}^{+1.0}_{-0.8} &   5.00 \\
 4702.99    & Mg I  &   4.330 & -0.380 &  {55.1}^{+0.8}_{-1.2} &   4.96 \\
 5172.68    & Mg I  &   2.710 & -0.450 & {173.9}^{+2.2}_{-2.5} &   5.06 \\
 5183.60    & Mg I  &   2.720 & -0.239 & {195.9}^{+1.8}_{-2.1} &   5.13 \\
 5528.40    & Mg I  &   4.340 & -0.498 &  {59.0}^{+1.5}_{-1.8} &   5.04 \\
 5711.09    & Mg I  &   4.340 & -1.724 &   {6.2}^{+1.4}_{-1.4} &   4.98 \\
 3961.52    & Al I  &   0.014 & -0.340 & {104.6}^{+1.6}_{-1.7} &   2.79 \\
 3905.52    & Si I  &   1.910 & -1.092 & {166.2}^{+5.3}_{-4.0} &   5.23 \\
 4102.94    & Si I  &   1.910 & -3.140 &  {60.4}^{+1.4}_{-1.8} &   4.96 \\
 7698.96    & K I   &   0.000 & -0.168 &  {15.3}^{+1.1}_{-1.2} &   2.19 \\
 4226.73    & Ca I  &   0.000 &  0.244 & {149.9}^{+1.4}_{-1.5} &   3.25 \\
 4283.01    & Ca I  &   1.886 & -0.224 &  {31.6}^{+2.1}_{-1.5} &   3.44 \\
 4318.65    & Ca I  &   1.890 & -0.210 &  {28.6}^{+1.3}_{-1.0} &   3.35 \\
 4425.44    & Ca I  &   1.879 & -0.358 &  {25.5}^{+1.1}_{-1.3} &   3.40 \\
 4434.96    & Ca I  &   1.886 & -0.010 &  {41.9}^{+1.3}_{-1.3} &   3.39 \\
 4435.69    & Ca I  &   1.890 & -0.519 &  {18.1}^{+1.6}_{-1.3} &   3.38 \\
 4454.78    & Ca I  &   1.898 &  0.260 &  {50.8}^{+1.0}_{-0.8} &   3.29 \\
 4455.89    & Ca I  &   1.899 & -0.530 &  {17.6}^{+1.1}_{-0.9} &   3.38 \\
 5588.76    & Ca I  &   2.520 &  0.210 &  {24.5}^{+1.5}_{-1.3} &   3.43 \\
 5594.47    & Ca I  &   2.523 &  0.097 &  {16.5}^{+1.8}_{-1.4} &   3.33 \\
 5598.49    & Ca I  &   2.521 & -0.087 &  {11.3}^{+1.5}_{-1.4} &   3.33 \\
 5857.45    & Ca I  &   2.930 &  0.230 &  {11.0}^{+1.7}_{-1.3} &   3.46 \\
 6102.72    & Ca I  &   1.880 & -0.790 &  {17.4}^{+2.1}_{-1.5} &   3.44 \\
 6122.22    & Ca I  &   1.890 & -0.315 &  {34.9}^{+1.3}_{-1.3} &   3.37 \\
 6162.17    & Ca I  &   1.900 & -0.089 &  {45.7}^{+1.6}_{-1.2} &   3.34 \\
 6169.56    & Ca I  &   2.526 & -0.478 &   {5.2}^{+1.2}_{-1.1} &   3.32 \\
 6439.07    & Ca I  &   2.520 & 0.470  &  {31.0}^{+1.3}_{-1.3} &   3.25 \\
 4246.00    & Sc I  & (synth) &        &                       &  -0.02 \\
 4323.00    & Sc I  & (synth) &        &                       &  -0.03 \\
 4415.00    & Sc I  & (synth) &        &                       &  -0.01 \\
 5030.00    & Sc I  & (synth) &        &                       &  -0.13 \\
 5526.00    & Sc I  & (synth) &        &                       &  -0.15 \\
 5656.00    & Sc I  & (synth) &        &                       &   0.04 \\
 4313.00    & Sc II & (synth) &        &                       &   0.23 \\
 3759.29    & Ti II &   0.610 &  0.280 & {142.6}^{+2.4}_{-2.4} &   1.88 \\
 3761.32    & Ti II &   0.570 &  0.180 & {142.4}^{+2.1}_{-2.6} &   1.92 \\
 3913.46    & Ti II &   1.120 & -0.420 &  {91.1}^{+2.6}_{-2.1} &   1.89 \\
 4012.40    & Ti II &   0.574 & -1.750 &  {63.0}^{+1.6}_{-1.4} &   1.80 \\
 4025.12    & Ti II &   0.607 & -1.980 &  {43.1}^{+1.4}_{-1.3} &   1.68 \\
 4028.34    & Ti II &   1.890 & -0.960 &  {18.8}^{+1.7}_{-2.1} &   1.71 \\
 4053.83    & Ti II &   1.893 & -1.210 &  {12.4}^{+1.2}_{-1.1} &   1.74 \\
 4161.53    & Ti II &   1.084 & -2.160 &  {16.7}^{+1.1}_{-1.1} &   1.83 \\
 4163.63    & Ti II &   2.590 & -0.400 &  {17.2}^{+1.2}_{-1.6} &   1.91 \\
 4290.22    & Ti II &   1.160 & -0.930 &  {81.4}^{+2.9}_{-2.0} &   1.97 \\
 4330.72    & Ti II &   1.180 & -2.060 &  {16.3}^{+1.2}_{-1.2} &   1.79 \\
 4394.06    & Ti II &   1.221 & -1.780 &  {23.2}^{+1.1}_{-0.9} &   1.74 \\
 4395.03    & Ti II &   1.080 & -0.540 &  {90.9}^{+0.9}_{-1.1} &   1.63 \\
 4399.77    & Ti II &   1.240 & -1.190 &  {53.7}^{+1.1}_{-1.0} &   1.75 \\
 4417.71    & Ti II &   1.165 & -1.190 &  {56.6}^{+1.0}_{-0.9} &   1.70 \\
 4418.33    & Ti II &   1.240 & -1.970 &  {16.8}^{+1.4}_{-1.3} &   1.77 \\
 4441.73    & Ti II &   1.180 & -2.410 &  {12.9}^{+1.2}_{-1.3} &   2.00 \\
 4443.80    & Ti II &   1.080 & -0.720 &  {84.7}^{+1.0}_{-1.2} &   1.65 \\
 4444.55    & Ti II &   1.120 & -2.240 &  {13.1}^{+0.9}_{-1.0} &   1.76 \\
 4450.48    & Ti II &   1.080 & -1.520 &  {47.2}^{+1.2}_{-0.9} &   1.75 \\
 4464.45    & Ti II &   1.160 & -1.810 &  {26.4}^{+0.9}_{-0.8} &   1.75 \\
 4468.52    & Ti II &   1.131 & -0.600 &  {84.7}^{+0.9}_{-0.9} &   1.59 \\
 4470.85    & Ti II &   1.165 & -2.020 &  {14.7}^{+0.8}_{-1.0} &   1.65 \\
 4501.27    & Ti II &   1.116 & -0.770 &  {78.6}^{+0.9}_{-0.9} &   1.60 \\
 4529.48    & Ti II &   1.572 & -2.030 &  {15.4}^{+1.4}_{-0.9} &   2.18 \\
 4533.96    & Ti II &   1.240 & -0.530 &  {83.5}^{+1.2}_{-1.0} &   1.61 \\
 4563.77    & Ti II &   1.221 & -0.960 &  {72.8}^{+1.2}_{-0.9} &   1.79 \\
 4571.97    & Ti II &   1.572 & -0.320 &  {72.7}^{+1.0}_{-1.0} &   1.58 \\
 4589.91    & Ti II &   1.237 & -1.790 &  {30.5}^{+0.8}_{-0.8} &   1.89 \\
 4708.66    & Ti II &   1.240 & -2.340 &   {8.8}^{+1.0}_{-0.8} &   1.78 \\
 4779.98    & Ti II &   2.048 & -1.370 &  {10.2}^{+1.3}_{-0.9} &   1.85 \\
 4805.09    & Ti II &   2.061 & -1.100 &  {15.9}^{+0.7}_{-0.8} &   1.82 \\
 5129.16    & Ti II &   1.890 & -1.240 &  {13.7}^{+2.1}_{-1.9} &   1.63 \\
 5185.90    & Ti II &   1.890 & -1.490 &   {8.6}^{+1.4}_{-1.6} &   1.65 \\
 5188.69    & Ti II &   1.580 & -1.050 &  {43.6}^{+1.8}_{-1.5} &   1.71 \\
 5336.79    & Ti II &   1.580 & -1.590 &  {19.0}^{+2.5}_{-2.0} &   1.74 \\
 4035.62    & V II  &   1.793 & -0.767 &  {13.3}^{+2.3}_{-3.1} &   1.10 \\
 4254.33    & Cr I  &   0.000 & -0.114 &  {87.2}^{+1.1}_{-1.1} &   1.90 \\
 4274.80    & Cr I  &   0.000 & -0.220 &  {84.2}^{+1.0}_{-0.9} &   1.92 \\
 4289.72    & Cr I  &   0.000 & -0.370 &  {76.1}^{+1.1}_{-1.2} &   1.86 \\
 4600.75    & Cr I  &   1.004 & -1.260 &   {8.7}^{+1.2}_{-1.3} &   2.36 \\
 4626.19    & Cr I  &   0.968 & -1.320 &   {7.0}^{+1.2}_{-1.0} &   2.28 \\
 4646.15    & Cr I  &   1.030 & -0.740 &  {16.7}^{+1.1}_{-1.1} &   2.19 \\
 4652.16    & Cr I  &   1.004 & -1.030 &  {11.2}^{+0.8}_{-1.1} &   2.25 \\
 5206.04    & Cr I  &   0.940 & 0.020  &  {58.0}^{+1.9}_{-1.5} &   2.04 \\
 5296.69    & Cr I  &   0.980 & -1.360 &   {9.9}^{+1.9}_{-1.8} &   2.40 \\
 5298.28    & Cr I  &   0.980 & -1.140 &  {12.7}^{+1.8}_{-2.0} &   2.30 \\
 5345.80    & Cr I  &   1.000 & -0.950 &  {15.7}^{+2.2}_{-1.9} &   2.24 \\
 5348.31    & Cr I  &   1.000 & -1.210 &   {9.3}^{+1.9}_{-1.6} &   2.24 \\
 5409.77    & Cr I  &   1.030 & -0.670 &  {20.9}^{+1.5}_{-2.1} &   2.14 \\
 4558.59    & Cr II &   4.070 & -0.656 &   {8.4}^{+1.0}_{-0.8} &   2.60 \\
 4588.14    & Cr II &   4.070 & -0.826 &   {6.6}^{+1.4}_{-1.2} &   2.65 \\
 4030.75    & Mn I  &   0.000 & -0.480 &  {93.4}^{+1.5}_{-1.5} &   1.81 \\
 4033.06    & Mn I  &   0.000 & -0.618 &  {80.6}^{+1.5}_{-1.2} &   1.61 \\
 4034.48    & Mn I  &   0.000 & -0.811 &  {72.1}^{+1.5}_{-1.8} &   1.58 \\
 4041.36    & Mn I  &   2.114 & 0.285  &  {15.6}^{+1.7}_{-1.7} &   1.82 \\
 4783.00    & Mn I  & (synth) &        &                       &   1.76 \\
 4823.53    & Mn I  &   2.319 & 0.144  &   {9.8}^{+1.2}_{-1.2} &   1.82 \\
 3689.46    & Fe I  &   2.940 & -0.168 &  {58.8}^{+3.6}_{-3.2} &   4.68 \\
 3753.61    & Fe I  &   2.180 & -0.890 &  {60.0}^{+2.0}_{-2.7} &   4.45 \\
 3765.54    & Fe I  &   3.240 & 0.482  &  {61.0}^{+2.3}_{-2.0} &   4.38 \\
 3786.68    & Fe I  &   1.010 & -2.185 &  {78.0}^{+2.5}_{-1.9} &   4.75 \\
 3805.34    & Fe I  &   3.300 & 0.313  &  {50.8}^{+1.9}_{-2.0} &   4.33 \\
 3839.26    & Fe I  &   3.047 & -0.330 &  {36.5}^{+1.8}_{-1.9} &   4.33 \\
 3841.05    & Fe I  &   1.610 & -0.044 & {117.7}^{+1.4}_{-2.0} &   4.34 \\
 3846.80    & Fe I  &   3.251 & -0.020 &  {43.6}^{+2.0}_{-2.2} &   4.42 \\
 3850.82    & Fe I  &   0.990 & -1.745 & {101.1}^{+2.2}_{-1.5} &   4.86 \\
 3852.57    & Fe I  &   2.176 & -1.180 &  {50.1}^{+1.7}_{-2.0} &   4.42 \\
 3863.74    & Fe I  &   2.692 & -1.430 &  {25.7}^{+2.2}_{-2.4} &   4.75 \\
 3867.22    & Fe I  &   3.017 & -0.450 &  {32.7}^{+2.1}_{-2.0} &   4.32 \\
 3885.51    & Fe I  &   2.424 & -1.090 &  {33.3}^{+1.8}_{-1.8} &   4.25 \\
 3887.05    & Fe I  &   0.910 & -1.140 & {121.1}^{+2.0}_{-2.0} &   4.61 \\
 3902.95    & Fe I  &   1.560 & -0.442 & {112.0}^{+1.5}_{-1.7} &   4.49 \\
 3917.18    & Fe I  &   0.990 & -2.155 &  {81.2}^{+1.6}_{-1.6} &   4.65 \\
 3940.88    & Fe I  &   0.958 & -2.600 &  {63.1}^{+1.8}_{-1.9} &   4.58 \\
 3949.95    & Fe I  &   2.180 & -1.251 &  {51.6}^{+1.6}_{-1.8} &   4.47 \\
 3977.74    & Fe I  &   2.198 & -1.120 &  {54.1}^{+2.2}_{-2.4} &   4.41 \\
 4001.66    & Fe I  &   2.176 & -1.900 &  {17.7}^{+1.1}_{-1.6} &   4.33 \\
 4005.24    & Fe I  &   1.557 & -0.583 & {110.6}^{+1.3}_{-1.2} &   4.49 \\
 4007.27    & Fe I  &   2.759 & -1.280 &  {18.2}^{+1.4}_{-1.4} &   4.43 \\
 4014.53    & Fe I  &   3.047 & -0.590 &  {44.2}^{+1.5}_{-1.4} &   4.68 \\
 4021.87    & Fe I  &   2.759 & -0.730 &  {39.5}^{+1.4}_{-1.3} &   4.37 \\
 4032.63    & Fe I  &   1.485 & -2.380 &  {29.9}^{+1.5}_{-1.6} &   4.27 \\
 4044.61    & Fe I  &   2.832 & -1.220 &  {17.0}^{+1.4}_{-1.2} &   4.41 \\
 4058.22    & Fe I  &   3.211 & -1.110 &  {15.5}^{+1.7}_{-1.5} &   4.70 \\
 4062.44    & Fe I  &   2.845 & -0.860 &  {29.1}^{+1.3}_{-1.3} &   4.37 \\
 4067.27    & Fe I  &   2.559 & -1.419 &  {21.2}^{+1.1}_{-1.6} &   4.39 \\
 4067.98    & Fe I  &   3.211 & -0.470 &  {31.1}^{+1.6}_{-1.6} &   4.47 \\
 4070.77    & Fe I  &   3.241 & -0.790 &  {17.5}^{+1.2}_{-1.3} &   4.48 \\
 4073.76    & Fe I  &   3.266 & -0.900 &  {12.2}^{+1.4}_{-1.3} &   4.43 \\
 4076.63    & Fe I  &   3.210 & -0.370 &  {28.2}^{+1.2}_{-1.7} &   4.29 \\
 4095.97    & Fe I  &   2.588 & -1.480 &  {14.3}^{+1.0}_{-1.2} &   4.27 \\
 4109.80    & Fe I  &   2.845 & -0.940 &  {28.3}^{+1.4}_{-1.4} &   4.42 \\
 4114.44    & Fe I  &   2.831 & -1.303 &  {15.2}^{+1.5}_{-1.5} &   4.41 \\
 4120.21    & Fe I  &   2.990 & -1.270 &  {14.8}^{+1.3}_{-1.8} &   4.55 \\
 4121.80    & Fe I  &   2.832 & -1.450 &  {12.9}^{+1.1}_{-1.2} &   4.48 \\
 4132.06    & Fe I  &   1.608 & -0.675 & {115.5}^{+1.4}_{-1.1} &   4.66 \\
 4132.90    & Fe I  &   2.845 & -1.010 &  {27.9}^{+1.3}_{-1.2} &   4.47 \\
 4134.68    & Fe I  &   2.830 & -0.649 &  {43.4}^{+1.2}_{-1.1} &   4.42 \\
 4137.00    & Fe I  &   3.415 & -0.450 &  {13.5}^{+1.2}_{-1.2} &   4.19 \\
 4139.93    & Fe I  &   0.990 & -3.629 &  {16.8}^{+1.1}_{-1.5} &   4.55 \\
 4143.41    & Fe I  &   3.047 & -0.200 &  {49.8}^{+1.3}_{-1.1} &   4.36 \\
 4143.87    & Fe I  &   1.557 & -0.511 & {116.0}^{+1.5}_{-1.0} &   4.43 \\
 4147.67    & Fe I  &   1.480 & -2.071 &  {58.6}^{+1.2}_{-1.1} &   4.49 \\
 4152.17    & Fe I  &   0.958 & -3.232 &  {36.3}^{+1.0}_{-1.2} &   4.57 \\
 4153.90    & Fe I  &   3.397 & -0.320 &  {28.0}^{+1.3}_{-1.2} &   4.45 \\
 4154.50    & Fe I  &   2.830 & -0.688 &  {38.5}^{+1.3}_{-1.4} &   4.35 \\
 4154.81    & Fe I  &   3.368 & -0.400 &  {26.1}^{+1.2}_{-1.0} &   4.44 \\
 4156.80    & Fe I  &   2.830 & -0.808 &  {38.1}^{+1.3}_{-1.3} &   4.46 \\
 4157.78    & Fe I  &   3.420 & -0.403 &  {22.4}^{+1.1}_{-1.4} &   4.42 \\
 4158.79    & Fe I  &   3.430 & -0.670 &  {12.7}^{+1.1}_{-1.5} &   4.39 \\
 4174.91    & Fe I  &   0.910 & -2.938 &  {56.7}^{+8.3}_{-6.1} &   4.60 \\
 4175.64    & Fe I  &   2.845 & -0.827 &  {35.3}^{+1.3}_{-1.3} &   4.44 \\
 4181.76    & Fe I  &   2.830 & -0.371 &  {55.3}^{+1.3}_{-1.3} &   4.36 \\
 4182.38    & Fe I  &   3.020 & -1.180 &  {10.4}^{+1.3}_{-1.4} &   4.31 \\
 4184.89    & Fe I  &   2.830 & -0.869 &  {28.0}^{+1.2}_{-1.3} &   4.30 \\
 4187.04    & Fe I  &   2.449 & -0.514 &  {68.3}^{+1.0}_{-1.0} &   4.32 \\
 4187.80    & Fe I  &   2.420 & -0.510 &  {73.4}^{+1.1}_{-0.9} &   4.39 \\
 4191.43    & Fe I  &   2.470 & -0.666 &  {60.5}^{+1.1}_{-1.1} &   4.32 \\
 4195.33    & Fe I  &   3.330 & -0.492 &  {27.6}^{+1.3}_{-1.2} &   4.51 \\
 4196.21    & Fe I  &   3.397 & -0.700 &   {9.5}^{+1.1}_{-1.2} &   4.23 \\
 4199.10    & Fe I  &   3.047 & 0.156  &  {64.8}^{+1.3}_{-1.2} &   4.29 \\
 4202.03    & Fe I  &   1.485 & -0.689 & {110.2}^{+1.1}_{-1.1} &   4.33 \\
 4216.18    & Fe I  &   0.000 & -3.357 &  {91.5}^{+1.5}_{-1.0} &   4.65 \\
 4217.55    & Fe I  &   3.430 & -0.484 &  {17.5}^{+1.1}_{-1.3} &   4.36 \\
 4222.21    & Fe I  &   2.449 & -0.914 &  {49.2}^{+1.1}_{-1.1} &   4.31 \\
 4227.43    & Fe I  &   3.332 & 0.266  &  {60.1}^{+1.5}_{-1.3} &   4.42 \\
 4233.60    & Fe I  &   2.482 & -0.579 &  {67.4}^{+1.2}_{-1.4} &   4.38 \\
 4238.81    & Fe I  &   3.400 & -0.233 &  {31.9}^{+1.3}_{-1.0} &   4.42 \\
 4247.43    & Fe I  &   3.368 & -0.240 &  {32.4}^{+1.6}_{-1.4} &   4.40 \\
 4250.12    & Fe I  &   2.469 & -0.380 &  {76.3}^{+1.2}_{-1.2} &   4.36 \\
 4250.79    & Fe I  &   1.557 & -0.713 & {105.0}^{+1.0}_{-1.0} &   4.28 \\
 4260.47    & Fe I  &   2.400 & 0.077  &  {97.3}^{+1.1}_{-1.0} &   4.33 \\
 4271.15    & Fe I  &   2.449 & -0.337 &  {82.4}^{+1.7}_{-1.2} &   4.43 \\
 4282.40    & Fe I  &   2.176 & -0.779 &  {70.5}^{+1.1}_{-1.2} &   4.25 \\
 4337.05    & Fe I  &   1.557 & -1.695 &  {73.2}^{+1.0}_{-1.2} &   4.43 \\
 4352.73    & Fe I  &   2.223 & -1.290 &  {53.9}^{+1.0}_{-1.2} &   4.45 \\
 4375.93    & Fe I  &   0.000 & -3.005 & {104.1}^{+1.0}_{-1.0} &   4.51 \\
 4404.75    & Fe I  &   1.557 & -0.147 & {133.9}^{+0.9}_{-1.1} &   4.28 \\
 4407.71    & Fe I  &   2.176 & -1.970 &  {27.3}^{+1.0}_{-1.1} &   4.55 \\
 4415.12    & Fe I  &   1.608 & -0.621 & {113.8}^{+0.9}_{-1.0} &   4.36 \\
 4422.57    & Fe I  &   2.845 & -1.110 &  {23.8}^{+1.1}_{-1.0} &   4.41 \\
 4427.31    & Fe I  &   0.050 & -2.924 & {105.4}^{+0.9}_{-1.0} &   4.49 \\
 4430.61    & Fe I  &   2.223 & -1.659 &  {34.2}^{+1.2}_{-1.1} &   4.43 \\
 4442.34    & Fe I  &   2.198 & -1.228 &  {59.5}^{+1.2}_{-1.1} &   4.44 \\
 4443.19    & Fe I  &   2.858 & -1.043 &  {21.9}^{+1.2}_{-1.1} &   4.31 \\
 4447.72    & Fe I  &   2.223 & -1.339 &  {51.1}^{+0.9}_{-0.9} &   4.42 \\
 4454.38    & Fe I  &   2.832 & -1.300 &  {16.6}^{+0.9}_{-1.0} &   4.38 \\
 4459.12    & Fe I  &   2.176 & -1.279 &  {63.0}^{+1.0}_{-1.2} &   4.52 \\
 4461.65    & Fe I  &   0.090 & -3.194 &  {96.1}^{+0.9}_{-0.9} &   4.57 \\
 4466.55    & Fe I  &   2.830 & -0.600 &  {59.7}^{+1.1}_{-1.0} &   4.58 \\
 4476.02    & Fe I  &   2.845 & -0.820 &  {45.9}^{+1.1}_{-1.0} &   4.56 \\
 4484.22    & Fe I  &   3.603 & -0.860 &  {12.4}^{+1.0}_{-1.2} &   4.71 \\
 4489.74    & Fe I  &   0.121 & -3.899 &  {58.4}^{+1.1}_{-1.2} &   4.50 \\
 4494.56    & Fe I  &   2.198 & -1.143 &  {64.7}^{+1.3}_{-0.9} &   4.44 \\
 4528.61    & Fe I  &   2.176 & -0.822 &  {79.2}^{+1.0}_{-1.1} &   4.38 \\
 4531.15    & Fe I  &   1.480 & -2.101 &  {60.5}^{+1.1}_{-1.4} &   4.42 \\
 4592.65    & Fe I  &   1.560 & -2.462 &  {40.5}^{+1.2}_{-0.9} &   4.50 \\
 4602.94    & Fe I  &   1.485 & -2.208 &  {57.6}^{+1.2}_{-1.0} &   4.46 \\
 4630.12    & Fe I  &   2.280 & -2.587 &   {6.7}^{+1.0}_{-1.1} &   4.53 \\
 4632.91    & Fe I  &   1.608 & -2.913 &  {18.9}^{+1.4}_{-1.4} &   4.55 \\
 4647.43    & Fe I  &   2.950 & -1.351 &  {14.1}^{+0.9}_{-1.1} &   4.46 \\
 4678.85    & Fe I  &   3.603 & -0.830 &  {12.8}^{+1.0}_{-1.1} &   4.67 \\
 4691.41    & Fe I  &   2.990 & -1.520 &  {10.1}^{+0.9}_{-1.0} &   4.51 \\
 4707.27    & Fe I  &   3.241 & -1.080 &  {16.3}^{+3.2}_{-2.9} &   4.61 \\
 4710.28    & Fe I  &   3.018 & -1.610 &   {7.8}^{+1.1}_{-1.3} &   4.51 \\
 4733.59    & Fe I  &   1.490 & -2.988 &  {20.2}^{+0.9}_{-1.1} &   4.50 \\
 4736.77    & Fe I  &   3.211 & -0.752 &  {27.3}^{+0.8}_{-0.9} &   4.52 \\
 4786.81    & Fe I  &   3.000 & -1.606 &   {5.3}^{+0.9}_{-1.0} &   4.29 \\
 4789.65    & Fe I  &   3.530 & -0.957 &   {7.9}^{+1.4}_{-1.5} &   4.46 \\
 4871.32    & Fe I  &   2.870 & -0.362 &  {58.8}^{+1.6}_{-1.2} &   4.28 \\
 4872.14    & Fe I  &   2.882 & -0.567 &  {47.6}^{+1.1}_{-1.3} &   4.30 \\
 4890.76    & Fe I  &   2.876 & -0.394 &  {56.8}^{+1.4}_{-1.3} &   4.28 \\
 4891.49    & Fe I  &   2.852 & -0.111 &  {69.5}^{+1.2}_{-1.2} &   4.19 \\
 4903.31    & Fe I  &   2.880 & -0.926 &  {30.1}^{+1.1}_{-1.2} &   4.33 \\
 4918.99    & Fe I  &   2.845 & -0.342 &  {61.7}^{+1.2}_{-1.5} &   4.27 \\
 4920.50    & Fe I  &   2.832 & 0.068  &  {79.8}^{+1.1}_{-1.5} &   4.18 \\
 4924.77    & Fe I  &   2.280 & -2.114 &  {14.6}^{+1.2}_{-1.5} &   4.39 \\
 4938.81    & Fe I  &   2.875 & -1.077 &  {26.0}^{+1.1}_{-1.2} &   4.38 \\
 4939.69    & Fe I  &   0.859 & -3.252 &  {41.1}^{+0.9}_{-0.9} &   4.38 \\
 4946.39    & Fe I  &   3.368 & -1.170 &   {7.9}^{+1.2}_{-1.1} &   4.46 \\
 4966.09    & Fe I  &   3.330 & -0.871 &  {16.6}^{+4.0}_{-2.7} &   4.47 \\
 4994.13    & Fe I  &   0.920 & -2.969 &  {56.0}^{+2.0}_{-2.1} &   4.41 \\
 5001.87    & Fe I  &   3.880 & 0.050  &  {16.9}^{+2.9}_{-2.9} &   4.22 \\
 5006.12    & Fe I  &   2.830 & -0.615 &  {46.7}^{+2.9}_{-2.8} &   4.25 \\
 5012.07    & Fe I  &   0.859 & -2.642 &  {84.5}^{+2.5}_{-2.6} &   4.51 \\
 5041.07    & Fe I  &   0.958 & -3.090 &  {49.1}^{+3.2}_{-3.9} &   4.45 \\
 5041.76    & Fe I  &   1.485 & -2.200 &  {63.0}^{+4.6}_{-4.6} &   4.45 \\
 5049.82    & Fe I  &   2.280 & -1.355 &  {48.2}^{+2.0}_{-2.2} &   4.33 \\
 5051.63    & Fe I  &   0.920 & -2.764 &  {73.4}^{+2.5}_{-2.2} &   4.49 \\
 5068.77    & Fe I  &   2.940 & -1.041 &  {25.2}^{+2.4}_{-2.4} &   4.39 \\
 5074.75    & Fe I  &   4.220 & -0.200 &  {12.5}^{+2.5}_{-3.0} &   4.71 \\
 5079.22    & Fe I  &   2.200 & -2.105 &  {25.5}^{+3.9}_{-4.4} &   4.56 \\
 5079.74    & Fe I  &   0.990 & -3.245 &  {41.6}^{+2.0}_{-1.9} &   4.52 \\
 5083.34    & Fe I  &   0.960 & -2.842 &  {58.4}^{+1.7}_{-2.1} &   4.35 \\
 5098.70    & Fe I  &   2.176 & -2.030 &  {30.9}^{+2.1}_{-2.5} &   4.56 \\
 5110.41    & Fe I  &   0.000 & -3.760 &  {86.8}^{+2.0}_{-1.6} &   4.57 \\
 5123.72    & Fe I  &   1.010 & -3.058 &  {47.2}^{+1.5}_{-2.1} &   4.44 \\
 5125.12    & Fe I  &   4.220 & -0.140 &  {11.1}^{+2.1}_{-2.2} &   4.59 \\
 5127.36    & Fe I  &   0.920 & -3.249 &  {41.4}^{+2.3}_{-1.9} &   4.42 \\
 5131.47    & Fe I  &   2.220 & -2.515 &   {8.8}^{+1.4}_{-1.6} &   4.45 \\
 5133.69    & Fe I  &   4.178 & 0.140  &  {18.7}^{+2.1}_{-1.6} &   4.52 \\
 5141.74    & Fe I  &   2.420 & -2.238 &   {8.5}^{+2.1}_{-1.7} &   4.40 \\
 5142.93    & Fe I  &   0.958 & -3.080 &  {47.0}^{+1.6}_{-1.6} &   4.39 \\
 5150.84    & Fe I  &   0.990 & -3.037 &  {45.5}^{+1.7}_{-1.6} &   4.36 \\
 5151.91    & Fe I  &   1.010 & -3.321 &  {31.0}^{+1.5}_{-1.5} &   4.42 \\
 5162.27    & Fe I  &   4.178 & 0.020  &  {12.8}^{+1.5}_{-1.5} &   4.45 \\
 5166.28    & Fe I  &   0.000 & -4.123 &  {61.2}^{+2.4}_{-2.5} &   4.46 \\
 5171.60    & Fe I  &   1.490 & -1.721 &  {84.3}^{+2.0}_{-1.5} &   4.33 \\
 5191.45    & Fe I  &   3.040 & -0.551 &  {38.6}^{+1.7}_{-1.6} &   4.27 \\
 5192.34    & Fe I  &   3.000 & -0.421 &  {49.9}^{+1.6}_{-1.8} &   4.29 \\
 5194.94    & Fe I  &   1.560 & -2.021 &  {64.7}^{+1.8}_{-1.5} &   4.36 \\
 5198.71    & Fe I  &   2.220 & -2.091 &  {21.5}^{+1.9}_{-1.7} &   4.46 \\
 5202.34    & Fe I  &   2.180 & -1.871 &  {37.5}^{+1.9}_{-1.4} &   4.52 \\
 5216.27    & Fe I  &   1.610 & -2.082 &  {57.2}^{+1.5}_{-1.3} &   4.36 \\
 5217.39    & Fe I  &   3.210 & -1.162 &  {10.9}^{+1.6}_{-1.7} &   4.39 \\
 5225.53    & Fe I  &   0.110 & -4.755 &  {19.8}^{+2.1}_{-1.3} &   4.47 \\
 5232.94    & Fe I  &   2.940 & -0.057 &  {72.8}^{+1.9}_{-1.5} &   4.24 \\
 5247.05    & Fe I  &   0.087 & -4.946 &  {17.0}^{+1.5}_{-1.6} &   4.55 \\
 5250.21    & Fe I  &   0.121 & -4.938 &  {16.9}^{+1.8}_{-2.1} &   4.58 \\
 5250.65    & Fe I  &   2.198 & -2.180 &  {25.7}^{+1.7}_{-1.6} &   4.62 \\
 5254.96    & Fe I  &   0.110 & -4.764 &  {22.9}^{+1.9}_{-1.5} &   4.55 \\
 5263.31    & Fe I  &   3.266 & -0.879 &  {15.8}^{+1.9}_{-1.6} &   4.35 \\
 5266.56    & Fe I  &   3.000 & -0.385 &  {53.9}^{+1.8}_{-1.6} &   4.31 \\
 5269.54    & Fe I  &   0.860 & -1.333 & {144.8}^{+1.7}_{-1.8} &   4.40 \\
 5281.79    & Fe I  &   3.040 & -0.833 &  {29.5}^{+1.9}_{-2.1} &   4.37 \\
 5283.62    & Fe I  &   3.240 & -0.524 &  {33.8}^{+1.6}_{-1.6} &   4.39 \\
 5302.30    & Fe I  &   3.283 & -0.720 &  {19.8}^{+1.6}_{-1.5} &   4.32 \\
 5307.36    & Fe I  &   1.610 & -2.912 &  {15.3}^{+1.6}_{-1.4} &   4.34 \\
 5324.18    & Fe I  &   3.211 & -0.103 &  {51.7}^{+1.7}_{-1.5} &   4.24 \\
 5328.04    & Fe I  &   0.915 & -1.466 & {132.8}^{+1.6}_{-1.6} &   4.34 \\
 5328.53    & Fe I  &   1.557 & -1.850 &  {76.8}^{+1.1}_{-1.5} &   4.38 \\
 5332.90    & Fe I  &   1.550 & -2.776 &  {19.7}^{+1.6}_{-1.8} &   4.26 \\
 5339.93    & Fe I  &   3.270 & -0.720 &  {22.5}^{+1.5}_{-1.5} &   4.37 \\
 5364.87    & Fe I  &   4.450 & 0.228  &  {14.1}^{+2.7}_{-1.6} &   4.59 \\
 5367.47    & Fe I  &   4.420 & 0.443  &  {14.1}^{+1.8}_{-1.4} &   4.34 \\
 5369.96    & Fe I  &   4.370 & 0.536  &  {19.2}^{+1.6}_{-1.6} &   4.35 \\
 5371.49    & Fe I  &   0.960 & -1.644 & {124.4}^{+1.6}_{-1.7} &   4.38 \\
 5383.37    & Fe I  &   4.310 & 0.645  &  {24.4}^{+1.2}_{-1.6} &   4.30 \\
 5393.17    & Fe I  &   3.240 & -0.910 &  {23.4}^{+1.5}_{-1.6} &   4.54 \\
 5397.13    & Fe I  &   0.920 & -1.982 & {114.9}^{+1.9}_{-1.6} &   4.46 \\
 5405.77    & Fe I  &   0.990 & -1.852 & {113.9}^{+1.8}_{-1.4} &   4.39 \\
 5410.91    & Fe I  &   4.470 & 0.398  &  {10.4}^{+1.6}_{-1.7} &   4.29 \\
 5415.20    & Fe I  &   4.390 & 0.643  &  {19.9}^{+1.3}_{-2.1} &   4.28 \\
 5424.07    & Fe I  &   4.320 & 0.520  &  {27.5}^{+2.0}_{-1.2} &   4.50 \\
 5429.70    & Fe I  &   0.960 & -1.881 & {116.0}^{+1.8}_{-1.5} &   4.42 \\
 5434.52    & Fe I  &   1.010 & -2.126 &  {98.2}^{+1.5}_{-1.5} &   4.35 \\
 5446.92    & Fe I  &   0.990 & -1.910 & {113.8}^{+1.2}_{-1.8} &   4.43 \\
 5455.61    & Fe I  &   1.010 & -2.090 & {115.7}^{+1.3}_{-1.7} &   4.67 \\
 5497.52    & Fe I  &   1.010 & -2.825 &  {63.7}^{+1.5}_{-1.5} &   4.41 \\
 5501.47    & Fe I  &   0.960 & -3.046 &  {61.3}^{+1.8}_{-1.6} &   4.53 \\
 5506.78    & Fe I  &   0.990 & -2.789 &  {70.4}^{+1.7}_{-1.4} &   4.46 \\
 5569.62    & Fe I  &   3.417 & -0.540 &  {24.6}^{+1.4}_{-1.6} &   4.40 \\
 5572.84    & Fe I  &   3.397 & -0.275 &  {30.4}^{+1.2}_{-1.6} &   4.24 \\
 5576.09    & Fe I  &   3.430 & -1.000 &  {13.0}^{+1.3}_{-1.5} &   4.54 \\
 5586.76    & Fe I  &   3.370 & -0.144 &  {43.3}^{+1.5}_{-1.3} &   4.30 \\
 5615.64    & Fe I  &   3.332 & 0.050  &  {53.3}^{+1.4}_{-1.2} &   4.22 \\
 5624.54    & Fe I  &   3.417 & -0.755 &  {16.9}^{+1.4}_{-1.4} &   4.41 \\
 5701.54    & Fe I  &   2.560 & -2.143 &   {7.6}^{+1.3}_{-1.4} &   4.36 \\
 6065.48    & Fe I  &   2.610 & -1.410 &  {26.6}^{+1.2}_{-1.1} &   4.29 \\
 6136.61    & Fe I  &   2.450 & -1.410 &  {42.5}^{+1.3}_{-1.7} &   4.38 \\
 6137.69    & Fe I  &   2.590 & -1.346 &  {35.6}^{+1.2}_{-1.2} &   4.37 \\
 6191.56    & Fe I  &   2.430 & -1.416 &  {39.9}^{+1.4}_{-1.2} &   4.31 \\
 6213.43    & Fe I  &   2.220 & -2.481 &   {9.3}^{+1.2}_{-1.4} &   4.34 \\
 6219.28    & Fe I  &   2.200 & -2.448 &  {14.9}^{+1.4}_{-1.6} &   4.51 \\
 6230.72    & Fe I  &   2.560 & -1.276 &  {41.3}^{+1.1}_{-1.3} &   4.35 \\
 6246.32    & Fe I  &   3.600 & -0.877 &   {9.6}^{+1.3}_{-1.3} &   4.42 \\
 6252.56    & Fe I  &   2.404 & -1.687 &  {30.4}^{+1.1}_{-1.0} &   4.38 \\
 6254.26    & Fe I  &   2.279 & -2.443 &  {12.9}^{+1.1}_{-1.3} &   4.52 \\
 6265.13    & Fe I  &   2.180 & -2.540 &  {10.3}^{+0.9}_{-1.0} &   4.39 \\
 6335.33    & Fe I  &   2.198 & -2.180 &  {21.4}^{+1.4}_{-1.3} &   4.41 \\
 6336.84    & Fe I  &   3.686 & -1.050 &   {6.7}^{+1.1}_{-1.1} &   4.52 \\
 6393.60    & Fe I  &   2.430 & -1.576 &  {38.5}^{+1.3}_{-1.0} &   4.43 \\
 6400.00    & Fe I  &   3.603 & -0.290 &  {22.8}^{+1.2}_{-1.3} &   4.27 \\
 6411.65    & Fe I  &   3.654 & -0.595 &  {12.3}^{+1.4}_{-1.4} &   4.32 \\
 6421.35    & Fe I  &   2.280 & -2.014 &  {26.6}^{+1.3}_{-1.4} &   4.46 \\
 6430.85    & Fe I  &   2.180 & -1.946 &  {30.8}^{+1.0}_{-1.0} &   4.36 \\
 6494.98    & Fe I  &   2.400 & -1.239 &  {68.7}^{+1.2}_{-1.0} &   4.52 \\
 6592.91    & Fe I  &   2.730 & -1.473 &  {21.7}^{+1.5}_{-1.2} &   4.35 \\
 6677.99    & Fe I  &   2.690 & -1.418 &  {29.7}^{+1.3}_{-1.5} &   4.41 \\
 6978.85    & Fe I  &   2.480 & -2.452 &   {8.0}^{+1.2}_{-1.5} &   4.50 \\
 4178.86    & Fe II &   2.583 & -2.510 &  {41.1}^{+1.4}_{-1.5} &   4.47 \\
 4233.17    & Fe II &   2.583 & -1.970 &  {71.8}^{+2.1}_{-1.9} &   4.53 \\
 4416.82    & Fe II &   2.778 & -2.600 &  {27.2}^{+1.1}_{-1.2} &   4.46 \\
 4489.19    & Fe II &   2.828 & -2.970 &  {15.0}^{+1.1}_{-1.0} &   4.54 \\
 4491.41    & Fe II &   2.860 & -2.710 &  {21.7}^{+1.1}_{-1.2} &   4.52 \\
 4508.28    & Fe II &   2.856 & -2.580 &  {36.5}^{+1.1}_{-0.8} &   4.70 \\
 4515.34    & Fe II &   2.844 & -2.600 &  {28.3}^{+1.0}_{-0.9} &   4.54 \\
 4520.22    & Fe II &   2.810 & -2.600 &  {27.8}^{+1.1}_{-0.9} &   4.49 \\
 4522.63    & Fe II &   2.840 & -2.250 &  {48.5}^{+1.4}_{-1.2} &   4.57 \\
 4541.52    & Fe II &   2.856 & -3.050 &  {13.1}^{+0.9}_{-0.8} &   4.58 \\
 4555.89    & Fe II &   2.828 & -2.400 &  {36.1}^{+1.6}_{-1.7} &   4.47 \\
 4576.34    & Fe II &   2.840 & -2.950 &  {13.7}^{+0.9}_{-1.1} &   4.48 \\
 4583.84    & Fe II &   2.807 & -1.930 &  {64.8}^{+1.1}_{-0.9} &   4.50 \\
 4620.52    & Fe II &   2.830 & -3.210 &   {5.5}^{+0.8}_{-1.0} &   4.28 \\
 4731.44    & Fe II &   2.891 & -3.360 &  {10.3}^{+1.2}_{-0.9} &   4.79 \\
 4923.93    & Fe II &   2.891 & -1.320 &  {82.7}^{+1.4}_{-1.8} &   4.24 \\
 5018.45    & Fe II &   2.891 & -1.220 &  {94.6}^{+1.9}_{-2.7} &   4.36 \\
 5197.58    & Fe II &   3.230 & -2.220 &  {25.5}^{+2.0}_{-2.3} &   4.47 \\
 5234.63    & Fe II &   3.220 & -2.180 &  {29.3}^{+1.9}_{-1.6} &   4.49 \\
 5276.00    & Fe II &   3.200 & -2.010 &  {32.9}^{+1.9}_{-1.6} &   4.37 \\
 5284.08    & Fe II &   2.891 & -3.190 &   {8.6}^{+1.9}_{-1.6} &   4.46 \\
 5534.83    & Fe II &   3.245 & -2.930 &   {8.9}^{+1.4}_{-1.1} &   4.63 \\
 6456.38    & Fe II &   3.900 & -2.080 &   {7.9}^{+1.2}_{-1.5} &   4.44 \\
 5342.00    & Co I  & (synth) &        &                       &  <2.91 \\
 3566.37    & Ni I  &   0.420 & -0.251 & {122.0}^{+5.0}_{-3.1} &   3.31 \\
 3597.71    & Ni I  &   0.210 & -1.115 &  {95.1}^{+4.9}_{-5.9} &   3.26 \\
 3783.52    & Ni I  &   0.423 & -1.420 &  {81.6}^{+3.6}_{-3.0} &   3.07 \\
 3807.14    & Ni I  &   0.423 & -1.220 &  {95.4}^{+2.5}_{-2.2} &   3.23 \\
 3858.30    & Ni I  &   0.423 & -0.951 &  {97.9}^{+2.0}_{-1.5} &   2.98 \\
 4648.66    & Ni I  &   3.420 & -0.160 &   {5.3}^{+0.8}_{-1.1} &   3.07 \\
 4714.42    & Ni I  &   3.380 &  0.230 &  {14.8}^{+1.0}_{-1.0} &   3.12 \\
 5476.90    & Ni I  &   1.830 & -0.890 &  {46.9}^{+1.6}_{-1.8} &   2.95 \\
 6643.64    & Ni I  &   1.680 & -2.300 &   {8.1}^{+1.4}_{-1.2} &   3.11 \\
 5218.00    & Cu I  & (synth) &        &                       &  <1.36 \\
 4722.15    & Zn I  &   4.030 & -0.390 &   {4.9}^{+0.9}_{-1.2} &   1.52 \\
 4810.53    & Zn I  &   4.080 & -0.137 &   {7.4}^{+0.6}_{-0.8} &   1.50 \\
 4172.00    & Ga I  &   0.100 & -0.310 &                    <5 &  <0.74 \\
 7800.28    & Rb I  &   0.000 &  0.140 &                    <5 &  <1.26 \\
 4077.71    & Sr II &   0.000 &  0.150 &  {15.3}^{+1.3}_{-1.2} &  -3.21 \\
 4215.52    & Sr II &   0.000 & -0.180 &  {16.9}^{+1.6}_{-1.5} &  -2.87 \\
 4398.01    & Y II  & (synth) &        &                       & <-1.62 \\
 4161.20    & Zr II & (synth) &        &                       & <-1.29 \\
 3818.86    & Nb II & (synth) &        &                       & <-0.37 \\
 3864.10    & Mo I  & (synth) &        &                       & <-0.70 \\
 3801.01    & Sn I  & (synth) &        &                       &  <0.71 \\
 4554.00    & Ba II & (synth) &        &                       &  -3.62 \\
 4086.71    & La II & (synth) &        &                       & <-2.68 \\
 5274.00    & Ce II & (synth) &        &                       & <-1.25 \\
 4408.81    & Pr II & (synth) &        &                       & <-1.36 \\
 4706.54    & Nd II & (synth) &        &                       & <-1.91 \\
 4467.34    & Sm II & (synth) &        &                       & <-1.62 \\
 4205.04    & Eu II & (synth) &        &                       & <-2.98 \\
 4251.73    & Gd II & (synth) &        &                       & <-2.01 \\
 4752.53    & Tb II & (synth) &        &                       & <-2.28 \\
 4449.70    & Dy II & (synth) &        &                       & <-1.98 \\
 3996.51    & Tm II & (synth) &        &                       & <-1.98 \\
 3694.20    & Yb II & (synth) &        &                       & <-3.24 \\
 4093.15    & Hf II & (synth) &        &                       & <-1.48 \\
 3800.12    & Ir I  & (synth) &        &                       & <-0.45 \\
 4057.00    & Pb I  & (synth) &        &                       & <-0.83 \\
\enddata
\end{deluxetable*}
\end{longrotatetable}

\end{document}